\documentclass{aa} 
\usepackage{graphicx}
\usepackage{txfonts}
\usepackage{natbib}
\bibpunct{(}{)}{;}{a}{}{,}
\begin{document}
   \title{New light on the driving mechanism in roAp stars}

   \subtitle{Part I. Effects of metallicity}

   \author{S. Th\'eado \inst{1,} \inst{2}\and M.-A. Dupret \inst{3} \and A. Noels \inst{2} \and J. W. Ferguson \inst{4}}

   \offprints{S. Th\'eado}

   \institute{Laboratoire d'Astrophysique de Toulouse-Tarbes, Universit\'e de Toulouse, CNRS, 14 avenue Edouard Belin, 31400 Toulouse, France\\
                \email{stheado@ast.obs-mip.fr}
         \and
            Institut d'Astrophysique et de G\'eophysique de Li\`ege, All\'e du 6 Ao\^ut 17, 4000 Li\`ege, Belgium\\
         \and
           Observatoire de Paris, LESIA, CNRS UMR 8109, 5 place J. Janssen, 92195 Meudon, France\\
          \and
            Department of Physics, Wichita State University, Wichita, KS 67260-0032, USA
          \\
             }

   \date{Received September 15, 1996; accepted March 16, 1997}


  \abstract
   {Observations suggest that a relationship exists between the driving mechanism of roAp star pulsations and the heavy element distribution in these stars.}
   {We attempt to study the effects of local and global metallicity variations on the excitation mechanism of high order p-modes in A star models.}
   {We developed stellar evolutionary models to describe magnetic A stars with different global metallicity or local metal accumulation profiles. These models were computed with CLES (``Code Li\`egeois d'\'evolution stellaire''), and the stability of our models was assessed with the non-adiabatic oscillation code MAD.}
   {Our models reproduce the blue edge of the roAp star instability strip, but generate a red edge hotter than the observed one, regardless of metallicity. Surprisingly, we find that an increase in opacity inside the driving region can produce a lower amount of driving, which we refer to as the ``inverse $\kappa$-mechanism''.}
   {}
   

   \maketitle

\titlerunning{New light on the driving mechanism in roAp stars}
\authorrunning{Th\'eado et al.}

\section{Introduction : general context}
\subsection{The magnetic Ap stars}
The magnetic Ap stars are slowly rotating, chemically peculiar A and sometimes F type stars. They exhibit non-uniform distributions of chemical elements, both laterally across their surfaces and vertically with height in their atmospheres (see Sect. \ref{chemcomp}). They have strong organised magnetic fields that are predominantly dipolar,
although evidence of deviations from dipolar structure are
observed. The magnitudes of the fields range
typically from a few hundred to a few thousand Gauss \citep[e.g.][]{landstreet92,mathys97,hubrig04,hubrig05,wade00}. They can however be stronger as demonstrated by \citet{hubrig05} and \citet{ryabchikova06} who detected magnetic fields stronger than 20kG in cool Ap stars.

\subsection{The roAp stars}
The coolest subgroup of magnetic Ap stars is the Ap SrCrEu group ($\simeq$ 6400K-10000K). While most of these cool Ap stars do not show any oscillations, some exhibit single- or multi- period pulsations ranging from 4 to 21 min. These pulsating stars are called ``rapidly oscillating Ap stars'' (
roAp stars) as opposed to non-oscillating stars called noAp stars.

Both groups do not differ significantly in rotation and magnetic field strength. The physical processes that create either roAp or noAp stars are not yet understood, although observers attempt to identify observational differences between roAp and noAp to solve this puzzle. We discuss this more extensively in Sect. 2.

Until now, 35 roAp stars have been detected \citep{kurtz82,martinez91,kurtz94,martinez94,girish01,elkin05}. Their rapid pulsations are interpreted as high-overtone, low degree, non-radial p-modes.

\section{Observational context}

\subsection{Position in the HR diagram}
The roAp stars have effective temperatures of between 6400K and 8400K. They occupy mostly the main sequence part of the classical instability strip, in a similar way to $\delta$ Scuti stars. However, a few are cooler than $\delta$ Scuti stars: e.g. HD 213637 one of the coolest confirmed roAp stars ($\rm T_{eff} \simeq$ 6400K, \citet{kochukhov03}).

Comparing observational points to the models of \citet{schaller92} (with the solar initial composition of \citet{anders89} and Z=0.02), \citet{hubrig00} concluded that roAp stars are located on average slightly above the zero-age main-sequence, and significantly below the TAMS. This conclusion needs to be reconsidered however following the discovery of two apparently evolved roAp stars: HD 213637 \citep{kochukhov03} and HD 116114 \citep{elkin05} (see Sect. \ref{noap}).We demonstrate however in the following sections how the evolutionary stage of roAp stars depend strongly on their initial chemical composition and, in particular on their global metallicity.

   \begin{figure*}
   \centering
   \includegraphics[width=0.4\textwidth]{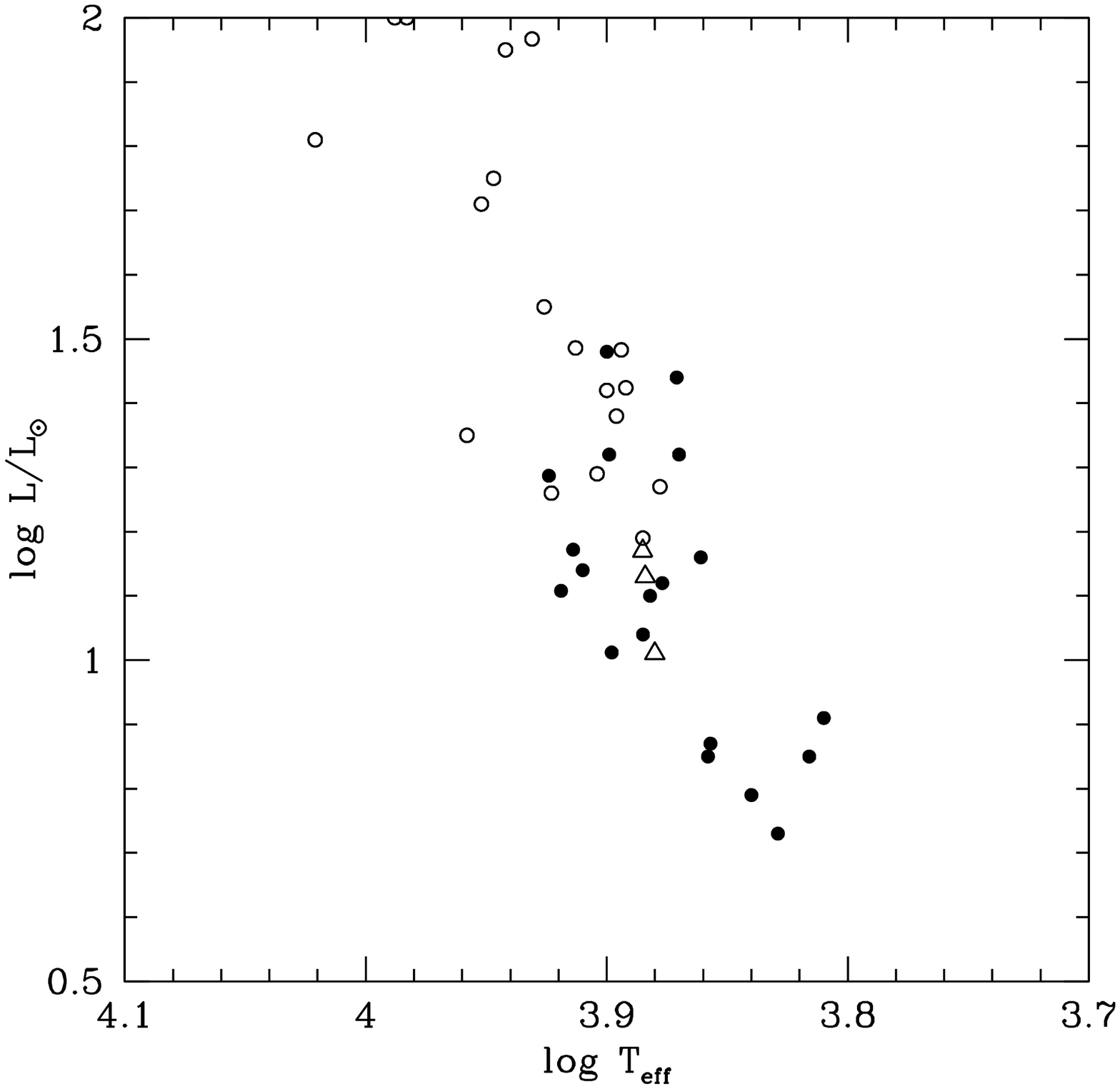}
   \includegraphics[width=0.4\textwidth]{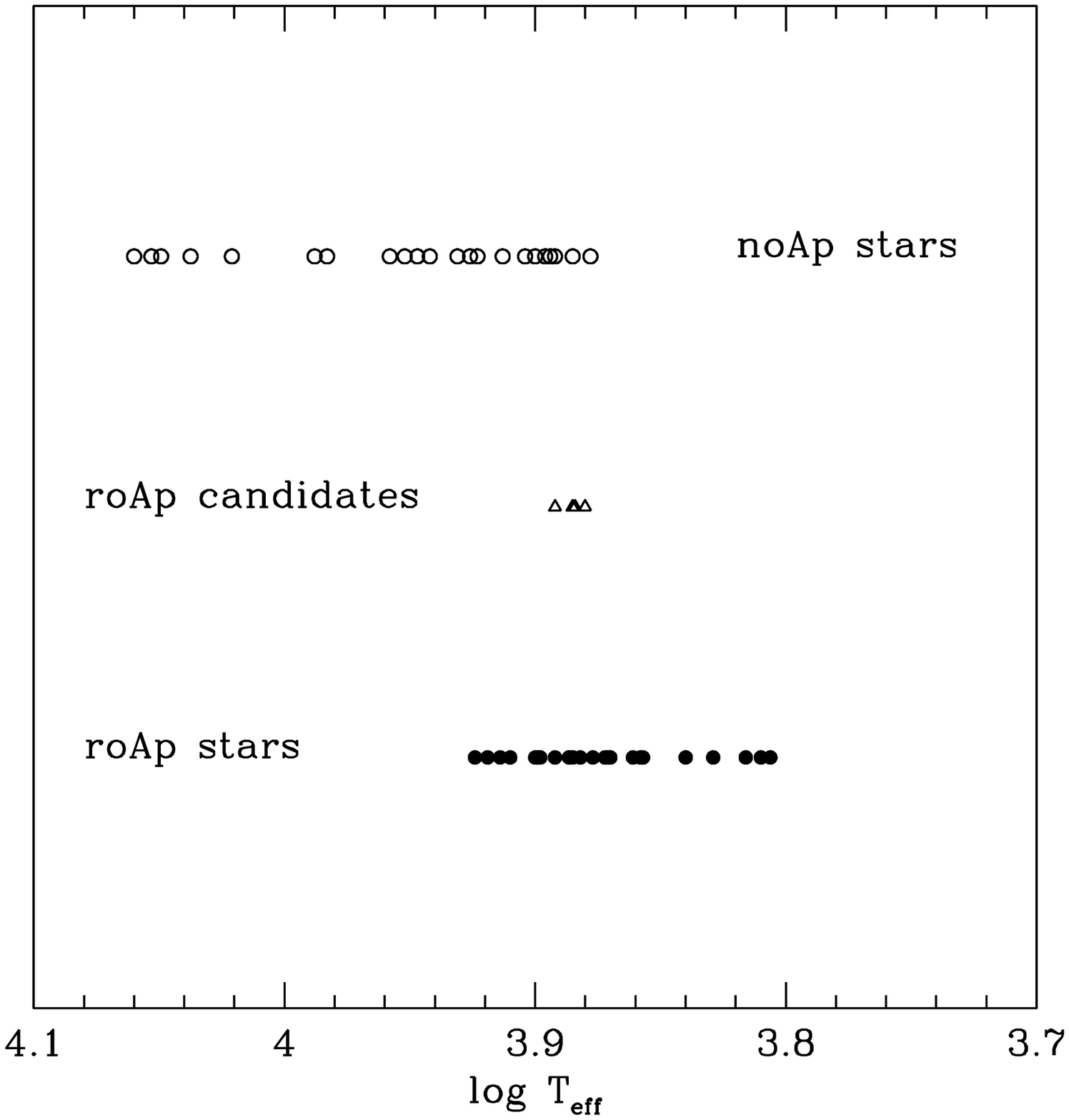}
   \caption{Observations of magnetic Ap stars. Left panel : Hertzsprung-Russell diagram. Solid circles represent 19 roAp stars, open circles, 18 noAp stars, and open triangles are 3 roAp star candidates. Right panel : effective temperatures of roAp and noAp stars and roAp stars candidates. Observations are taken from \citet{kochukhov06} (15 roAp stars + 3 candidates) and \citet{north97}.}
              \label{diaghr}
    \end{figure*}

\subsubsection{roAp versus noAp}
\label{noap}
Figure \ref{diaghr} displays the effective temperature and the luminosity of roAp stars, noAp stars, and roAp candidates. The left panel shows the HR diagram of stars for which both the luminosity and the effective temperature are available in the literature (19 roAp, 18 noAp, and 4 roAp candidates taken from \citet{kochukhov06} and \citet{north97} and the right panel shows the effective temperature of magnetic Ap stars for which only the effective temperature is known (23 roAp, 4 roAp candidates, and 24 noAp).

From the left and right panels of Fig. \ref{diaghr}, it is evident that roAp and noAp stars have a small overlap in effective temperature, the group of roAp stars being cooler than noAp stars. We also note the absence of cool non-pulsating Ap stars with effective temperatures lower than 7500 K, which suggests that all magnetic Ap stars up to a transition temperature of about 7500K-8100K may be affected by an excitation mechanism that is sufficiently strong to stimulate pulsations. 

Observations of the external parameters of roAp and noAp stars have suggested that noAp stars are more evolved and
 more luminous than roAp stars \citep{north97,hubrig00}.
However, as discussed
by \citet{hubrig00,cunha02,kurtz04}, this pattern could be due to observational biases. According to \citet{cunha02}, this bias could be related to the frequency of the
excited oscillations, which are expected to be lower in more evolved
and more luminous stars and possibly undetectable. As mentioned by \citet{hubrig00} and \citet{kurtz04},
most of the confirmed roAp stars were discovered by Martinez and Kurtz using photometric observations \citep[e.g.][]{kurtz82,martinez91,kurtz94,martinez94}. In their studies, they preferentially searched for stars with negative Str\"omgren
 $\delta c_1$ indices, because they are characteristic of strong chemical peculiarities in cool Ap stars and therefore provide good roAp star candidates. However, $\delta c_1$ increases with luminosity so that a peculiar evolved cool Ap star may show a normal $\delta c_1$. By focusing their surveys on stars with negative $\delta c_1$, Martinez and Kurtz limited their research to main-sequence (or close to main-sequence) stars, which introduced a selection effect into their discoveries. \citet{kurtz04} also remarked that the highspeed photometric searches of Martinez and Kurtz were inadequate for determining periods of up to 20-25 min, and may instead introduce a bias into the determination of the roAp frequency range upper limit.

Observations have indeed detected luminous and/or evolved roAp stars. From a comparison between the atmospheric parameters of the cool roAp star HD 213637 and evolutionary tracks computed by \citet{schaller92}, \citet{kochukhov03} concluded that it is either located close to the end of its main-sequence life or has already ascended the giant branch. \citet{elkin05} confirmed the existence of the luminous, evolved, lower frequency
roAp stars, with the discovery of a 21 minute pulsation in the cool
magnetic Ap star HD 116114 ($\rm T_{eff}=7413K$, $\rm \log (L/L_{\odot})=1.32$). One implication of this discovery is that there may be more luminous roAp stars with yet longer periods. A search is now being undertaken to discover more of these stars and study them in detail. If they are found in significant numbers, then the luminosity difference between the roAp and the noAp stars may disappear, leaving an unanswered question: why do some Ap stars pulsate and do not others ?

\subsection{Chemical composition}
\label{chemcomp}
\subsubsection{Surface Composition}
\label{surfacecomp}
The cool magnetic Ap stars are chemically peculiar stars, which exhibit
strongly anomalous, non-uniform element distributions. Their chemical peculiarities are distributed in patches about their surfaces, which are generally measured to contain underabundances in some light elements, such as C, O and sometimes N; almost solar to significantly low abundances of Fe and Ni (as well as other iron-peak element abundances); overabundances of Cr and Co; and significant
 overabundances of rare earth elements \citep{kupka96,gelbmann97,gelbmann98,ryabchikova97a,ryabchikova97b,ryabchikova00,ryabchikova05,gelbmann00}.
 An iron underabundance appears to be a common feature of the chemical composition of the atmospheres of roAp stars with effective temperatures of below 7000 K (HD 213637, \citet{kochukhov03}; HD 101065, \citet{cowley00}; HD 217522, \citet{gelbmann98}), while, for hotter stars, iron appears to be close to the solar abundance values \citep{kochukhov03,ryabchikova04}.

A Doppler imaging inversion technique has been developed \citep{piskunov93} and used \citep{kochukhov02,kochukhov04,lueftinger03} to reconstruct chemical maps of the surface of magnetic Ap stars from a series of high resolution spectral line profiles. \citet{kochukhov02} and \citet{kochukhov04} demonstrated that even if the surface distribution of some chemical species form symmetric patterns closely following the magnetic geometry, a multitude of chemical abundance structures in the magnetic Ap (and roAp) stars cannot be reduced to a system of spots and rings; they also showed that the distribution of some elements exhibits clear asymmetry with respect to the dipolar magnetic field geometry.

This suggests that phenomena other than the magnetic field interact with microscopic diffusion and play a substancial role in determining the geometry of abundance structures (e.g. rotation and mass loss).

\subsubsection{Vertical stratification}
\label{verticalstrat}

Abundances analyses derived from spectroscopic observations of magnetic Ap stars have discovered inconsistencies between abundances deduced from the weak and strong spectral lines and the lines of different ions. These anomalies provide clear evidence of prominent vertical abundance stratification of many chemical species, such as Ca,Cr, Fe, Na, Mg, Ba, Si, REE elements, \citep{bagnulo01,ryabchikova01,ryabchikova02,ryabchikova03,kochukhov03,kochukhov04,wade03}.

It appears to be widely accepted that iron (and most iron-peak elements) is normal to underabundant in magnetic Ap stars upper atmospheres, but numerous pieces of evidence highlight the increase in iron abundance deeper into the atmosphere  as in the roAp star HD 213637 \citep{kochukhov03}, the roAp star $\gamma$ Equ \citep{ryabchikova02}, and the Cr Ap star HD 204411 \citep{ryabchikova05}. For instance, the stratification model of \citet{ryabchikova02} for $\gamma$ Equ predicts that Ca, Cr, Fe, Ba, Si, and Na are concentrated in lower atmospheric layers, but are normal to underabundant in the upper layers with a transition between $\rm -1.5 < \log \tau_{5000} < -0.5$.

Globally, no obvious differences between the abundance pattern of Ap stars that pulsate or do not (or are candidate for pulsations) have been observed. However, an anomaly is detected in the REE abundances.
In roAp stars, abundances determined from lines of the third spectrum of various rare earth elements (notably Pr III and Nd III) appear to be overabundant from 1.5 dex and up to 2.5 dex relative to singly ionized spectra. This anomaly is observed in spectra of all investigated roAp stars \citep{weiss00,ryabchikova00,ryabchikova01,gelbmann00,cowley00,kochukhov03,ryabchikova04}, apart from HD 137909 and HD 116114,
which will be discussed later.

Similar analyses for non-pulsating Ap stars with otherwise similar properties did not find evidence of this imbalance or only marginally \citep{weiss00,ryabchikova04}. \citet{ryabchikova02} interpreted the roAp star anomaly as the result of a stratified Nd and Pr distribution. In the roAp star $\gamma$ Equ, they detected an accumulation of these elements above $\rm \log \tau_{5000}=-8$. \citet{mashonkina05} investigated the formation of Nd II and Nd III lines in the atmosphere of $\gamma$ Equ and HD 24712, and confirmed the detection by \citet{ryabchikova02} of enhanced Nd abundance layers, although for a different measurement of the transition at $\rm \log \tau_{5000}=-3.5$.

However \citet{ryabchikova02} did not find a REE anomaly in two luminous and evolved roAp stars: HD 137909 ($\rm T_{eff} = 7430K$ , $\rm \log (L/L_{\odot})= 1.44$) and HD 116114 ($\rm T_{eff}= 7413K$, $\rm \log (L/L_{\odot})= 1.32$) confirmed to be roAp stars \citep{hatzes04,elkin05}. This suggests that the true differences between roAp and noAp stars are still far from being understood.

\subsubsection{Metallicity of roAp stars}
Significant deviations in the chemical composition of CP stars from solar values are believed to be limited to surface layers. These anomalies are probably caused by a complex interplay between microscopic diffusion (including radiative levitation), mass loss, convection, turbulent mixing, and magnetic field. These complex effects are not well understood and are difficult to simulate in the context of 1D stellar evolution computations. Therefore, it is impossible to estimate the interior metal content from the observed surface abundance patterns.

According to observations from \citet[][Fig. 1]{gelbmann98}, \citet[][Fig.6]{kochukhov03} and \citet[][Fig. 5]{ryabchikova04}, the surface metallicity of the magnetic Ap stars is correlated with their effective temperature. The metallicity of stars with effective temperatures of between 6000K and 10000K increases with effective temperature. Iron is underabundant in stars cooler than 7000K (by up to one order of magnitude), close to solar in stars between 7000K and 8000K, and overabundant in stars hotter than 8000K. Since no roAp star hotter than 8700K was detected, this could imply that a relationship exists between the excitation mechanism and the heavy element distribution.

\section{Theoretical Context}

\subsection{Excitation mechanism}
Since the discovery of roAp stars, several mechanisms have been proposed and
tested extensively to excite their rapid oscillations: the direct influence of the Lorentz force \citep{dziembowski84,dziembowski85}, the magnetic overstability \citep{shibahashi83,cox84}, the stochastic excitation \citep[see][]{houdek99}, the $\kappa$-mechanism \citep{dolez82,matthews88,dolez88,dziembowski96,balmforth01}. For a review of all these processes we refer the reader to \citet{balmforth01} and \citet{cunha02}. In this paper, we discuss only the most plausible explanation: the excitation by the $\kappa$-mechanism.

As a result of their position in the HR diagram, close to the $\delta$ Scuti stars, it was first suggested that the oscillations of roAp stars might be driven by the same mechanism as $\delta$ Scuti modes, that is the $\kappa$-mechanism of HeII \citep[e.g.][]{kurtz90}. \citet{matthews88} also suggested that the roAp pulsations could be driven by the $\kappa$-mechanism acting in the SiIV ionization region, close to the HeII ionization region. While no accurate computations were carried out to test a possible excitation in the SiIV ionization zone, several studies tested excitation by the HeII $\kappa$-mechanism. Numerical computations, however, never supported this conjecture. As shown by \citet{dziembowski96} and confirmed by \citet{balmforth01}, only low frequency modes similar to those of $\delta$ Scuti stars can be excited in the helium second ionization region. Both sets of authors demonstrated that, in contrast, the HI ionization zone can exert significant driving for the high order p modes detected in roAp stars. Today the $\kappa$-mechanism operating in the HI ionization region is considered to be the most probable driving mechanism for roAp oscillations. As a result, the properties of the excited modes are expected to be strongly dependent on the chemical distribution, especially in the external layers of the star.

\subsection{Non-standard chemical transport processes}
Several non-standard processes are expected to be involved in the definition of the chemical element distribution in magnetic A stars. Numerous observations have been found to be consistent with the hypothesis that microscopic diffusion including radiative levitation, is a key process in these stars. The inhomogeneous and abnormal distribution of heavy elements observed at the surface of the magnetic Ap stars as well as the vertical stratification detected in their atmospheres (see section \ref{verticalstrat}) attest that element segregation effectively occurs in Ap stars, and that competing processes, if any, do not efficiently inhibit its effects.

Outflows of mass in the form of stellar winds are also expected to occur in Ap stars as they are observed in hotter and cooler main sequence stars. Unfortunately, no accurate measurement of mass loss-rates of A stars, only upper limits, have been available until now from observations ($\rm \dot{M} < 10^{-10}$ M$\rm _{\odot}.yr^{-1}$, \citet{lanz92,brown90}.
 The observed abundance anomalies in Ap and Am stars may also provide an indirect estimation of A-star mass-loss rates: several theoretical works demonstrate that mass-loss rates lower than $\rm 10^{-12}$ M$\rm _{\odot}.yr^{-1}$ (and sometimes as small as $\rm 10^{-15}$ M$\rm _{\odot}.yr^{-1}$ ) are needed to provide quantitative agreement between computed and observed abundances in chemically peculiar A stars \citep{michaud86,michaud83,babel91}.

Turbulent transport is likely to occur in stellar radiative interiors \citep[see][and references therein]{zahn93} due to various hydrodynamical instabilities. However, theoretical works dedicated to the magnetic field/turbulence (or convection) interaction and observations of abnormal abundances at the surface of magnetic Ap stars appear to exclude the presence of strong turbulent mixing (see next section) in these stars.

\subsubsection{Magnetic field effects}
\label{mfe}
The significant magnetic fields detected in magnetic Ap stars are expected to influence strongly their oscillation properties. In the deep interior, the magnetic field is unlikely to play an important role in the dynamics of the oscillations. In the outer layers, it is expected, however, to influence the properties of the oscillations both directly, by its effects on the mode geometry and frequencies and indirectly, by its effects on the chemical transport processes and the excitation mechanism.

The magnetic boundary layer influences basic properties of the pulsations, such as the oscillation frequencies and eigenfunctions \citep{dziembowski96,bigot00,cunha00,saio04,saio05,cunha07}. In particular, it can induce frequency shifts of between a few microhertz and a few tens of microhertz. \citet{saio05} showed that the direct effects of the magnetic field on the oscillations could stabilize the low radial order
 pulsations in roAp stars, and thus provided an explanation of the absence of these $\delta$ Scuti type modes in roAp stars otherwise predicted to be excited by the stellar evolution standard theory.

The magnetic field is expected to influence most chemical transport processes occurring in A stars. It can decelerate the microscopic diffusion but this effect is not expected to be significant in Ap stars \citep{theado05}. The magnetic field can also prevent stellar winds from flowing close to the magnetic equator \citep[see][and references therein]{theado05}. The main effects of the magnetic field, however, are related to convection and turbulence. As discussed previously \citep{gough66,moss69,shibahashi83,cox84,balmforth01}, a strong magnetic field can  freeze convective and turbulent motions in Ap stars in a significant part of the stars. This conjecture is supported by observations (described in Sect. \ref{verticalstrat}) which provide clear evidence of strong vertical abundance gradients in the atmosphere of magnetic Ap stars, indicating that the magnetic field inhibits convective and turbulent mixing in the outer stellar envelope.

Observations presented by \citet{kochukhov07} prompt a reconsideration of this conclusion.
They investigated the variations in the REE lines of roAp stars and discovered a change in the profile variability pattern with height in the atmospheres of all studied stars. They proposed that the line width modulation is a consequence of the periodic expansion and compression of turbulent layers in the upper atmospheres of roAp stars. This suggests that some cool Ap stars possess a turbulent zone possibly due to convective instability; this would only be true, however in the upper atmospheric layers (above $\rm \log \tau_{5000} \simeq -3.5$), since, in the intermediate atmosphere, significant vertical gradients of chemical composition preclude substantial mixing ($\rm -3.5 \le log \tau_{5000} \le -0.5$). This proposition is supported by theoretical diffusion models by \citet{leblanc04}, which demonstrate that convective mixing is needed to achieve a good agreement between stratification profiles inferred from computations and observations. These results suggest that the interaction between magnetic field and turbulent/convective motions could be more complex than previously anticipated.

\subsection{Modeling : status report}
Theorists have attempted to model the complex internal structure of magnetic Ap stars to explain, in particular, the properties of their oscillations. We now summarize the work completed to date.

Standard theoretical models that adequately describe roAp stars predict the excitation of $\delta$-Scuti modes but until now no high frequency modes were found to be unstable in these models \citep[e.g.][]{dolez82,dziembowski96,balmforth01}. As a result, alternative non-standard models were proposed to test the influence of various physical processes or structures such as a chromosphere, a magnetic field, helium settling, stellar winds, or mixing on the excitation mechanism of pulsations.

\citet{gautschy98} investigated the influence of an ad-hoc temperature inversion on the atmosphere of chemically homogeneous models. This inversion, which simulates the presence of a chromosphere enables roAp pulsations to be excited in models of A stars with effective temperatures lower than $\rm \log T_{eff}=3.86$. This theoretical blue edge for the instability strip is far cooler than observed. As emphasized by \citet{gautschy98}, there is also no observational evidence for the presence of a chromosphere in roAp stars.

Following works by \citet{gough66,moss69,shibahashi83,cox84} dedicated to the ability of magnetic fields to freeze convection, several authors \citep{balmforth01,cunha02} tested the stability of models with fully radiative envelope. They demonstrated that the suppression of convection in the external layers of the stars allows us to excite roAp pulsations in models with homogeneous composition. However, the theoretical instability strip derived by citet{cunha02} illustrated how these models do not account for the position of the roAp stars instability strip: the predicted theoretical red edge was too hot compared with observations.

\citet{theado05} presented the most accurate evolutionary models computed to date for roAp stars. They tested the influence of helium settling, weak stellar winds, and mild mixing on stellar models with convection suppressed in the external layers. The stability analyses of some of these models were presented in \citet{theado06}.
In models with He settling and convection suppressed, the external layers (including the H and He ionization regions) become rapidly poor in helium and enriched in hydrogen. As a result, excitation by the $\kappa$-mechanism is expected to be enhanced in the HI ionization region and be weaker in the HeII ionization zone. Stability analysis demonstrates that diffusion accelerates the damping of $\delta$ Scuti modes but the influence of He settling on the excitation of roAp modes is insignificant.
When a small wind (about $\rm 10^{-14}$ M$\rm _{\odot}.yr^{-1}$) competes with microscopic diffusion, an accumulation of helium occurs somewhere below the surface of the star, in the vicinity of the helium first ionization region. As expected, the He accumulation does not affect the excitation of roAp pulsations but favours the excitation of $\delta$ Scuti mode pulsations.
These models do not enable the observed instability strip to be reproduced. In particular, they fail to account for the excitation of pulsations in cool magnetic Ap stars, i.e. roAp stars observed in the lower right-hand-side corner of Fig. \ref{diaghr}.

\subsection{Theoretical challenges}
Until now no theoretical study has been able to account for the exact position of the roAp instability strip, in particular the excitation of roAp modes in magnetic Ap stars with effective temperatures lower than 7400K. On the other hand, the detection of both roAp and noAp stars in the upper part of the HR diagram and the absence of noAp stars with effective temperatures lower than 7550K both still present challenges to theorists. Abundance determinations in several roAp stars appear to imply that a relationship exists between the metallicity and the excitation mechanism. However, no study was dedicated to the influence of the heavy element distribution on the pulsations of magnetic Ap stars. When examining the most important contributors to the opacity in the H ionization zone, we also found, surprisingly, that, for surface conditions typical of roAp stars, a main contributor is iron, as can be seen in Figs. \ref{kappag1}, \ref{kappapz}, and \ref{kappapFe}.

The aim of this paper is to study the influence of metals on the excitation mechanism of roAp star pulsations and to assess whether these elements could help to resolve some open issues regarding roAp star properties.

\section{Basic non-adiabatic pulsation equations}
\label{eqsection}
As mentioned previously, roAp stars pulsate in low degree high order non-radial p-modes. Since the radial motions of a low degree high-radial-order-p mode are much larger than the horizontal motions, the results concerning the stability of radial pulsations are applicable to pulsations in roAp stars.
We recall the basic equations of linear non-adiabatic radial pulsations, since they are useful to the forthcoming discussion.
The additional terms originating in the magnetic field and rotation are
disregarded. The time-dependence of the eigenfunction is written
as: $\delta X(r,t)=\delta X(r)\exp(i\sigma t)$. We therefore have $\Im(\sigma)<0$
for unstable modes. The operator $\delta$ denotes Lagrangian perturbations.

\noindent The perturbed movement equation is given by:

\begin{equation}
\label{moveq}
\frac{{\rm d}\delta P}{{\rm d} r}\:=\:g\rho\left(4+\frac{\sigma^2 r}{g}\right)
\frac{\delta r}{r}\:.
\end{equation}

\noindent The continuity equation is:

\begin{equation}
\label{maseq}
\frac{\delta\rho}{\rho}+\frac{1}{r^2}\,\frac{{\rm d}(r^2\delta r)}{{\rm d}r}\:=0.
\end{equation}

\noindent The perturbed equation of energy conservation in the absence of nuclear reactions is:

\begin{equation}
\label{energeq}
i\:\sigma\:T\:\delta s\;=\;-\,\frac{{\rm d}\delta L}{{\rm d} m} \:.
\end{equation}

\noindent The perturbed diffusion equation in a full radiative zone is:

\begin{equation}
\label{dLeq}
\frac{\delta L}{L}\:=\:4\,\frac{\delta r}{r}\:+\:4\,\frac{\delta T}{T}
\:-\:\frac{\delta\kappa}{\kappa}\:+\:\frac{{\rm d}(\delta T/T)}{{\rm d}\ln T}\:.
\end{equation}
			
\noindent The work integral expression is:
\begin{eqnarray}
W(m) &=&  -\int_0^m \Im\left\{\frac{\delta\rho^*}{\rho}
\frac{\delta P}{\rho}\right\}{\rm d}m \nonumber\\&=&-\int_0^m(\Gamma_3-1)
\Im\left\{\frac{\delta\rho^*}{\rho}T\delta s\right\}\,{\rm d}m \nonumber\\
&=&\int_0^m\nabla_{\rm ad}\;\Im\left\{T\delta s^*
\frac{\delta P}{P}\right\}\,{\rm d}m\;.
\label{workeq}
\end{eqnarray}

\noindent The growth rate of the modes $\eta$ ($\delta X(r,t) = \delta X(r)\exp((i\Re(\sigma)+\eta)\,t)$)  is given by:
\begin{equation}
\eta = \frac{W(M)}{2\,\Re(\sigma)\,\int_0^M |\delta r|^2 {\rm d}m}\:.
\end{equation}

The integrant in Eq. (\ref{workeq}) can be simplified in regions where the local wavelength $\lambda = 2 \pi / k = 2 \pi c / \sigma$ is far shorter than the scale heights of equilibrium quantities.
Substituting Eq. (\ref{energeq}) into Eq. (\ref{workeq}), retaining only the term ${\rm d}(\delta T/T)/{\rm d}\ln T$ in
Eq. (\ref{dLeq}), and neglecting everywhere the eigenfunctions compared with their derivatives:

\begin{eqnarray}\label{eqdamp}
d W &\simeq& \frac{L}{|{\rm d}\ln T/{\rm d}r|}~
\Re\left\{\frac{1}{\sigma}\frac{{\delta T}}{T}^*
\frac{{\rm d}^2(\delta T/T)}{{\rm d}r^2}\right\}\:{\rm d}r\nonumber\\
&\simeq&-\;
\frac{\Re(\sigma) L}{c^2|{\rm d}\ln T/{\rm d}r|}~
\left|\frac{{\delta T}}{T}\right|^2\:{\rm d}r\;.
\end{eqnarray}

Hence the regions of short wavelength oscillations have a damping effect, generally refered to as ``radiative damping''.

\noindent Finally, the perturbed equations of state are:

\begin{eqnarray}
\label{eteq1}
\frac{\delta T}{T}&=&\frac{\delta s}{c_p}\:+\:\nabla_{\rm ad}\frac{\delta P}{P}\:,\\
\frac{\delta P}{P}&=P_T&\frac{\delta s}{c_v}\:+\:\Gamma_1\frac{\delta \rho}{\rho}\:.
\label{eteq2}
\end{eqnarray}

\section{Results}
We computed stellar models to describe A stars using the stellar evolution code Cl\'es (Code Li\'egeois d'Evolution Stellaire). For a complete description of the code, we refer the reader to \citet{scuflaire08}. The metal mixture used in these computations was the solar mixture presented by \citet{asplund05} (hereafter AGS05). The opacity tables are those of OPAL96 \citep{iglesias96} completed at low temperature with tables based upon calculations from \citet{ferguson05}\footnotemark \footnotetext{The opacity tables adequate for low temperatures are taken from the website http://webs.wichita.edu/physics/opacity/}. All these tables are computed using the composition taken from \citet{asplund05}. We use the OPAL01 equation of state \citep{rogers02}. As outer boundary conditions, Kurucz atmospheres \citep{kurucz98} are connected to the interior at an optical depth $\tau$ equal to 1. Following works by \citet{balmforth01} and \citet{theado05}, we computed models with and without convection in the envelope.

\subsection{Models with solar metallicity}

\subsubsection{Instability strip}

\begin{figure}[h]
\resizebox{\hsize}{!}{\includegraphics{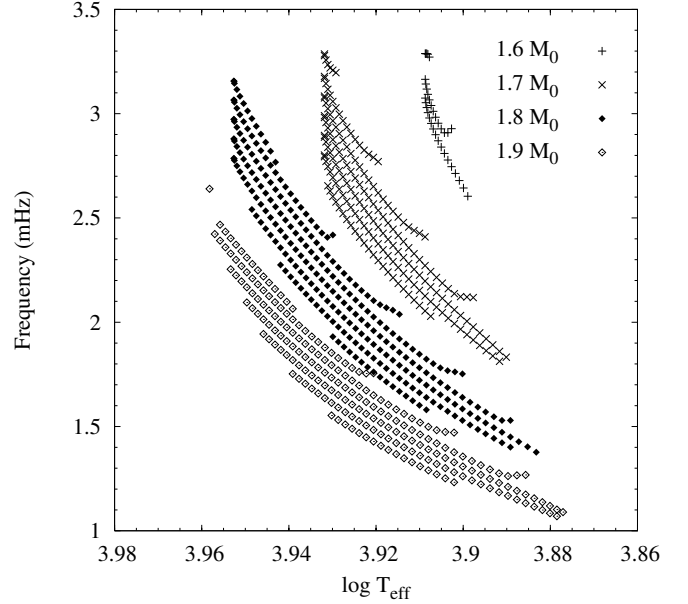}}
\caption{\label{freq}
Frequencies of the unstable modes predicted by our models as a function of $\log T_{\rm eff}$.
The different symbols correspond to different masses from
1.6 M$_\odot$ (cross) to 1.9 M$_\odot$ (diamond). The differences observed in the excited frequency ranges are mainly due to the
different dynamical times: $\tau_{\rm dyn} = (R^3/GM)^{1/2}$ of the models.
}
\end{figure}

We first consider models with initial mass fractions of hydrogen X=0.71 and solar metallicity according to AGS05 (i.e. Z/X=0.0165). These models were computed with a full radiative envelope. The stability of these models has been computed using the non-adiabatic pulsation code MAD \citep{dupret02a}. We restricted our research to the roap frequency domain without investigating the possible presence of excited $\delta$-Scuti modes (see Sect. \ref{mfe}). The left middle panel of Fig.~\ref{isz} shows the evolution tracks of our models and the calculated theoretical instability strip of roAp stars. Figure \ref{freq} displays, for four masses, the frequencies of the unstable roAp type modes along the evolutionary tracks. 
Figure  \ref{freq} shows that, although agreement is good down to a luminosity of about $\log (L/L_\odot) =0.9$, our models fail to reproduce the position of the red edge of the instability strip.
This is not a surprise since until now no evolutionary computation has been able to account for the excitation of roAp modes in stars with effective temperatures lower than $7400~K$. All computed models, whether of standard homogeneous composition \citep{balmforth01,cunha02} or with a stratified distribution induced by He settling and stellar winds \citep{theado05,theado06}, failed to shift the red limits towards cooler temperatures.

\subsubsection{Driving and damping regions}

In the present study, we neglected the coupling of
oscillations with the magnetic field; for the radiative transfer we use the diffusion approximation
below the photosphere and in the atmosphere we adopt the perturbed treatment of the thermal aspects 
proposed by \citep{dupret02b}. 
In this simplified framework, it is easier to identify and interpret
the main features of the driving mechanism in roAp stars.

\begin{figure}[t]
\resizebox{\hsize}{!}{\includegraphics{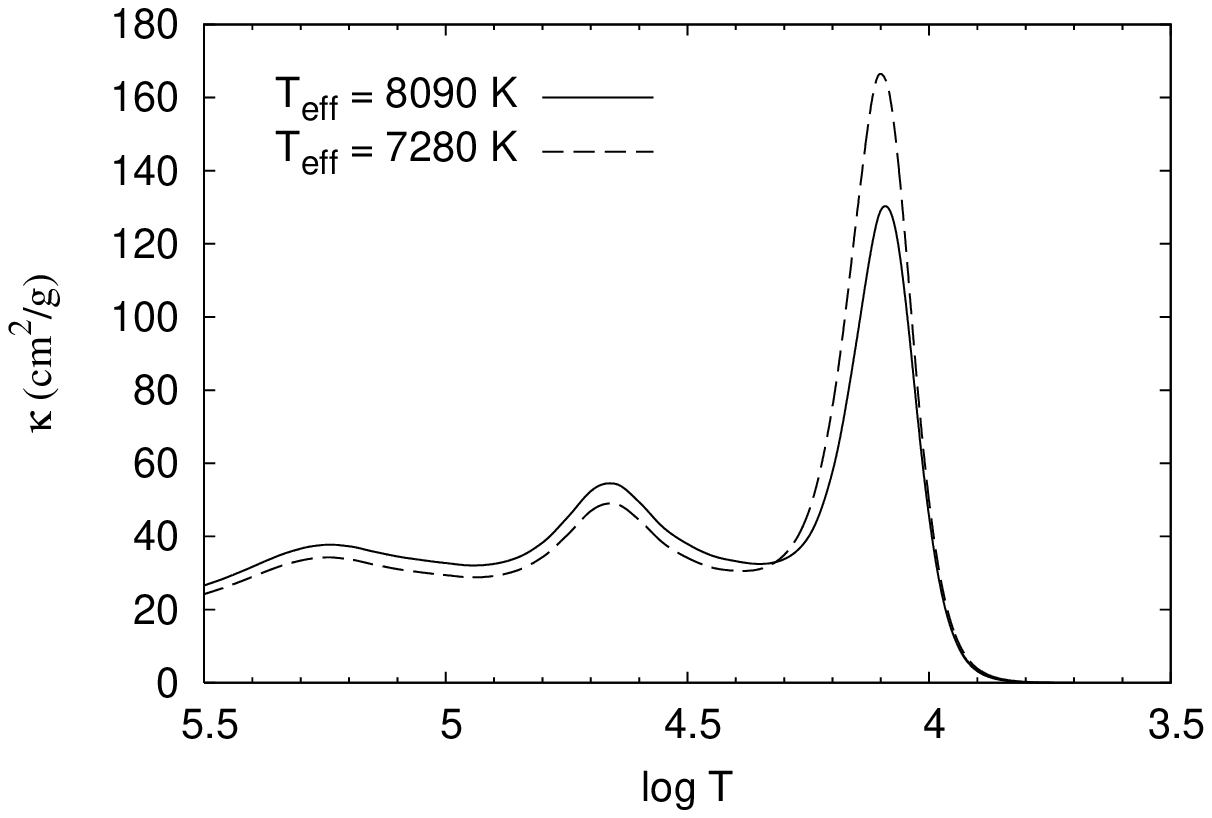}}
\caption{\label{kap}Opacities (cm$^2$/g) in two 1.6 M$_\odot$ models with different $T_{\rm eff}$.}
\resizebox{\hsize}{!}{\includegraphics{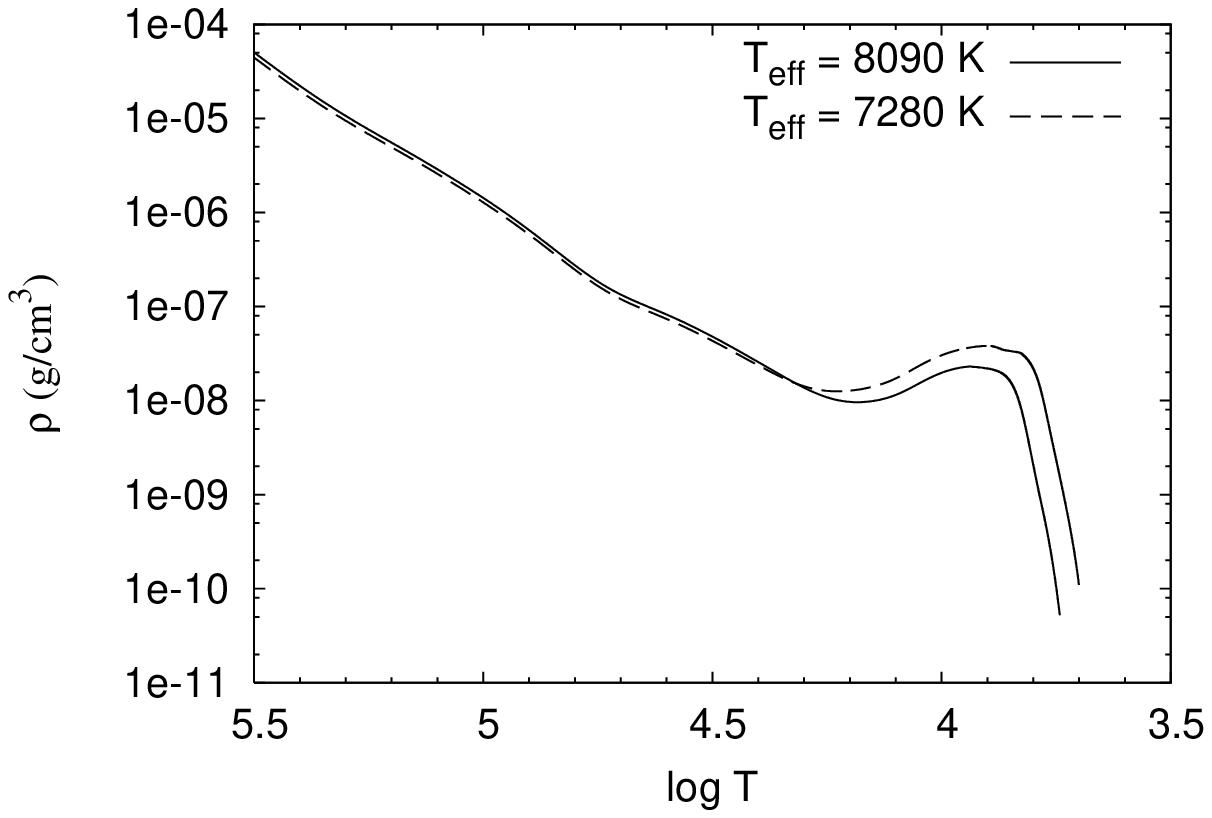}}
\caption{\label{rho}Densities of two 1.6 M$_\odot$ models with different $T_{\rm eff}$.}
\end{figure}

\begin{figure}[t]
\resizebox{\hsize}{!}{\includegraphics[width=17.5pc]{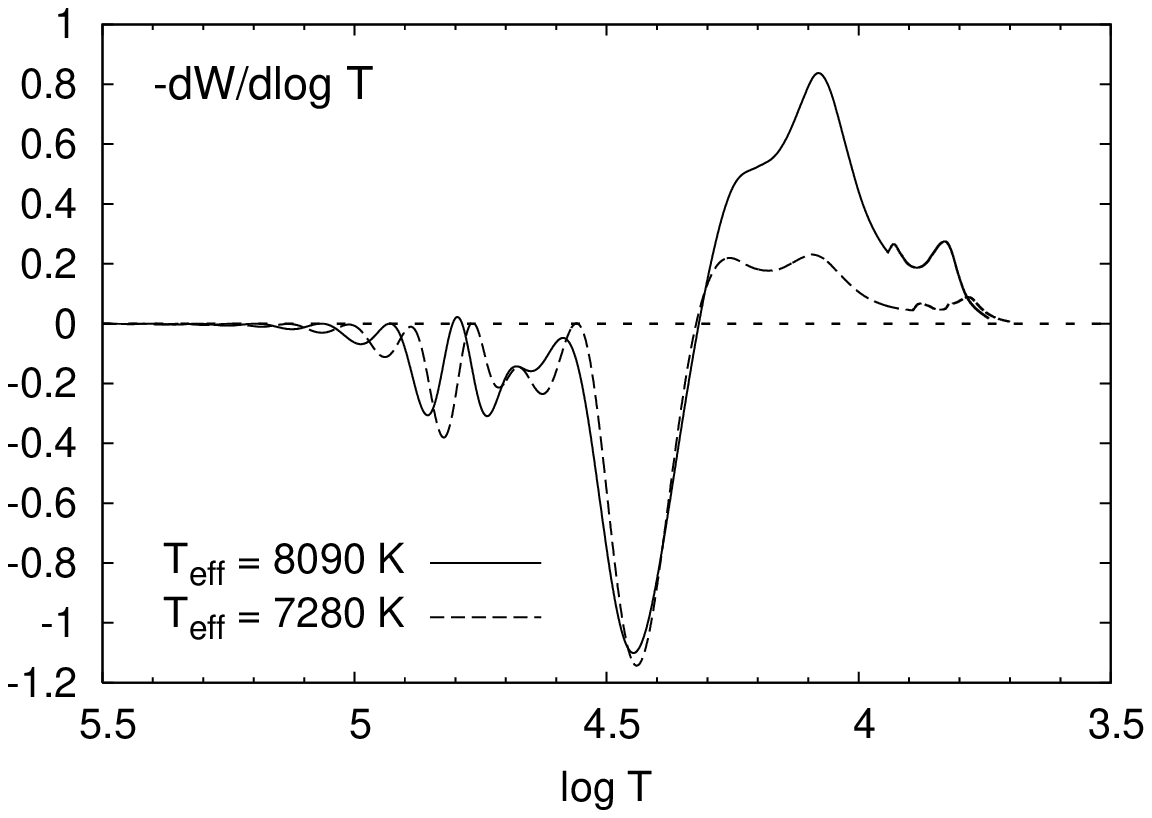}}
\caption{\label{work-red}Differential work $-{\rm d}W/{\rm d}\log T$ for the radial mode p$_{32}$ in two 1.6 M$_\odot$ models with different $T_{\rm eff}$.}
\resizebox{\hsize}{!}{\includegraphics[width=17.5pc]{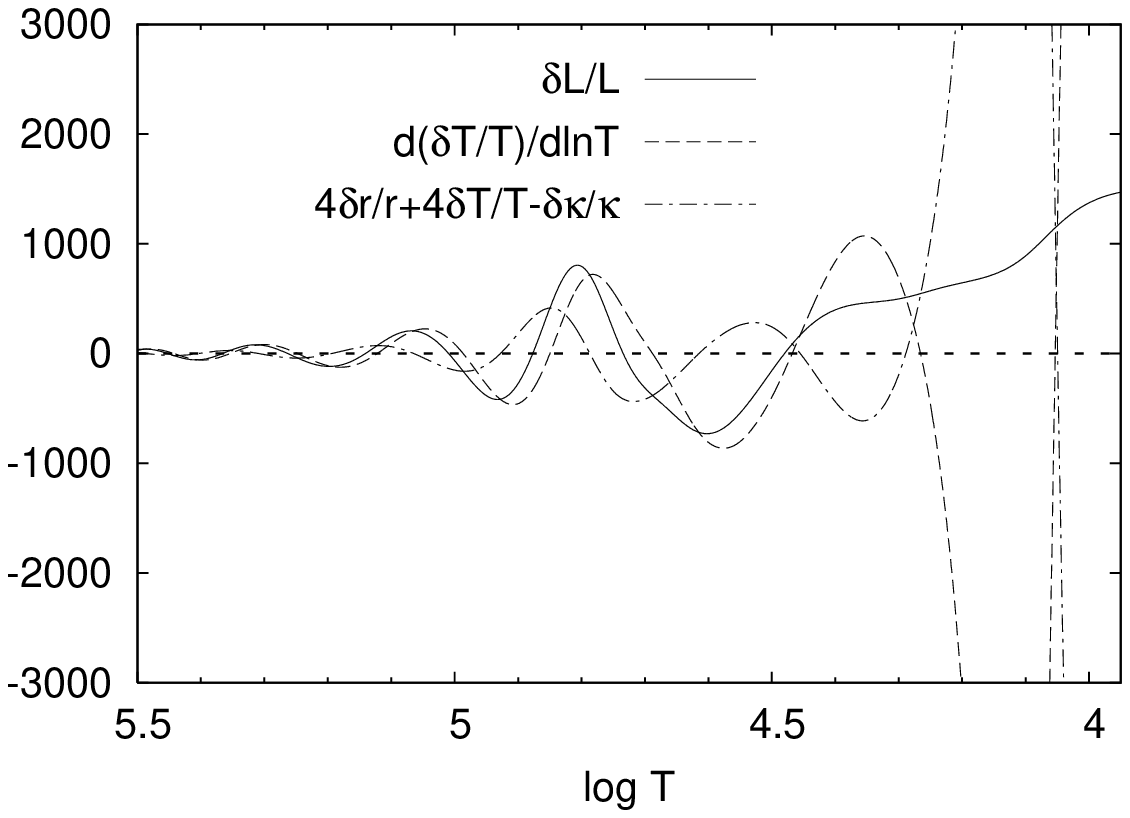}}
\caption{\label{dL-comp}Contributions to $\Re\{\delta L/L\}$  for the radial mode p$_{32}$  (see details in the text),
model with $T_{\rm eff}=8090$~K.}
\end{figure}

Figure \ref{kap} presents the opacities of two 1.6 M$_\odot$, [Fe/H]=0.00 models inside (full curve) and outside (dashed curve) of the theoretical instability strip. In this section we compare in detail the results obtained for these two models, represented by stars in the left middle panel of Fig. \ref{isz}.
Figure \ref{rho} shows the density distribution for these two models. We note the density inversions resulting from the large opacity bumps.
Figure \ref{work-red} presents the differential work $-{\rm d}W/{\rm d}\log T$ for the radial mode p$_{32}$ ($\nu=3$ mHz for the hot model, and $\nu=1.8$ mHz for the cold model). The regions where the differential work is positive (negative) have a driving (damping) effect on the oscillations. Damping occurs between $\log T = 5$ and $\log T = 4.3$ and driving occurs around the $\log T \simeq 4.1$ opacity bump.

In Fig. \ref{dL-comp}, we indicate the contribution of the different terms in Eq. \ref{dLeq}.
The term ${\rm d}(\delta T/T)/{\rm d}\ln T$ is given by the dashed line, the other contributions to $\delta L/L$ are shown by the point-dashed line, and the
solid line indicates the sum. Two distinct regions can be distinguished. From the center to
$\log T \simeq 4.5$, the term ${\rm d}(\delta T/T)/{\rm d}\ln T$ dominates
because the eigenfunctions have many nodes there. Substituting this dominating term in
Eq.~\ref{workeq} shows the well known radiative damping mechanism occurring in regions
of short-wavelength oscillations. We note that this radiative damping depends
on the equilibrium temperature gradient, but not significantly on the fluctuations of the opacity.
We now consider the region between $\log T \simeq 4.25$ and the photosphere.
Because of the strong opacity bump, $\delta T/T$ increases dramatically in this region. This can also be seen in Fig. \ref{ds-dt}, which indicates $|\delta T/T|$ and $|\delta s/c_P|$.

Since the pressure is determined by the movement equation
${\rm d}\delta P/{\rm d} r\:=\:g\rho(4+\sigma^2 r/g) (\delta r /r)$,
it cannot follow the enormous temperature variations. For $|\delta P/P|<<|\delta T/T|$, we obtain
from the perturbed equation of state (Eqs. 8 and 9):
$\delta s/c_p \approx \delta T/T \approx -\delta\rho/\rho$ in the H$_{\rm I}$ opacity bump
(see Fig. \ref{ds-dt}).
The heat exchanges are therefore extremely large, and significant amounts of energy can be absorbed from or transferred to the modes.
The precise result of this driving or damping depends on the respective phases of the eigenfunctions. Since $\delta\rho/\rho$  is more or less in opposite phase with  $\delta s/c_p$ and $\delta T/T$, it does not help to know if we have either driving or damping. It is more helpful to consider
the phase differences with respect to $\delta P$. From Eq. \ref{workeq}, we can see that to achieve a driving of the modes ($0^\circ<\phi(\delta P)-\phi(\delta s)<180^\circ$),
heat must be received (${\rm d}\delta Q/{\rm d}t>0$) when $\delta P>0$.
This appears to be the case for the radial mode p$_{32}$ considered here and it is therefore excited, but we will see in Sect. \ref{nodesect} that the driving of a modes strongly depends on the location of its last node.

The driving mechanism of roAp stars is therefore not identical to the weakly nonadiabatic $\kappa$-mechanism (as in other classical pulsators).
In the latter case, the fact that $\partial\ln\kappa/\partial\ln P|_s$ and therefore
$|\delta\kappa/\kappa|$ increase outwards implies automatically that ${\rm d}\delta L/{\rm d}m<0$ (Eq. 4),
so heat is gained during contraction and transformed into mechanical work.
In the H$_{\rm I}$ opacity bump of roAp stars, the opacity variations are counterbalanced by the temperature
gradient variations, and the result of positive or negative heat exchanges
and mechanical work depend on the modes.

\subsubsection{What determines the location of the red edge of the roAps instability strip ?}
\label{rededge}

\begin{figure}[t]
\resizebox{\hsize}{!}{\includegraphics{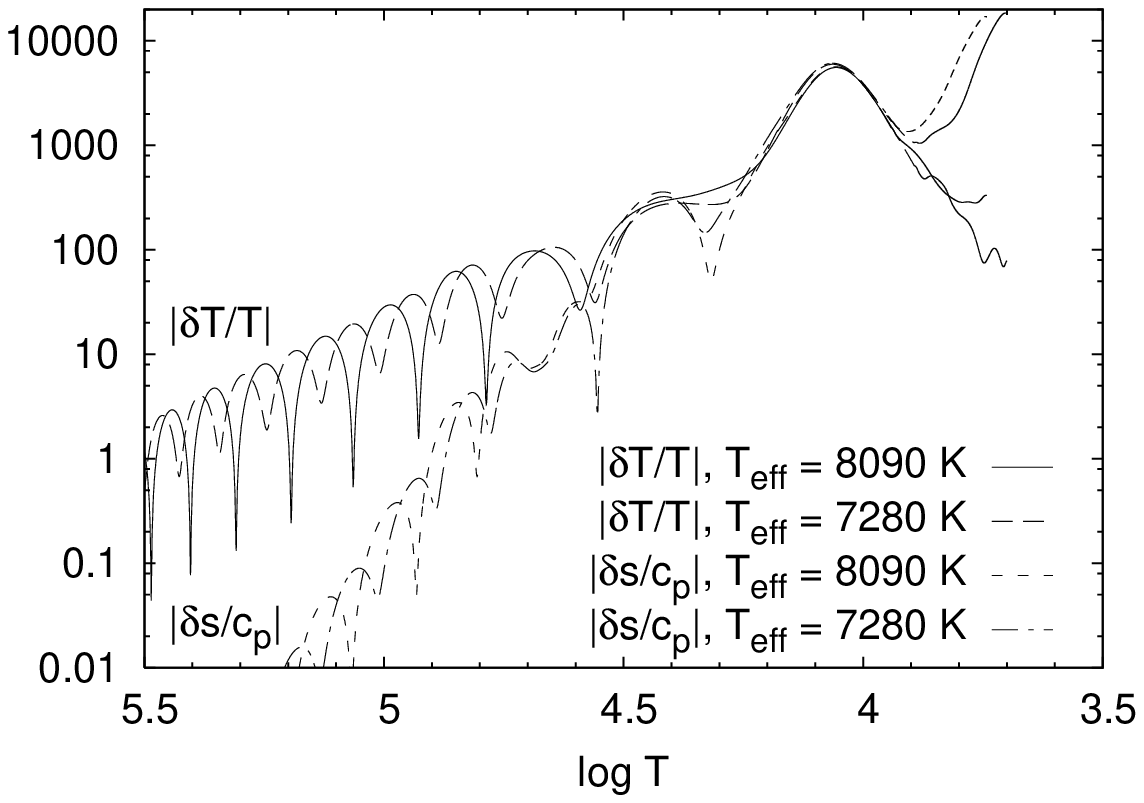}}
\caption{\label{ds-dt}$|\delta T/T|$ and $|\delta s/c_p|$ for the radial mode p$_{32}$ in two 1.6 M$_\odot$ models with different $T_{\rm eff}$.}
\resizebox{\hsize}{!}{\includegraphics{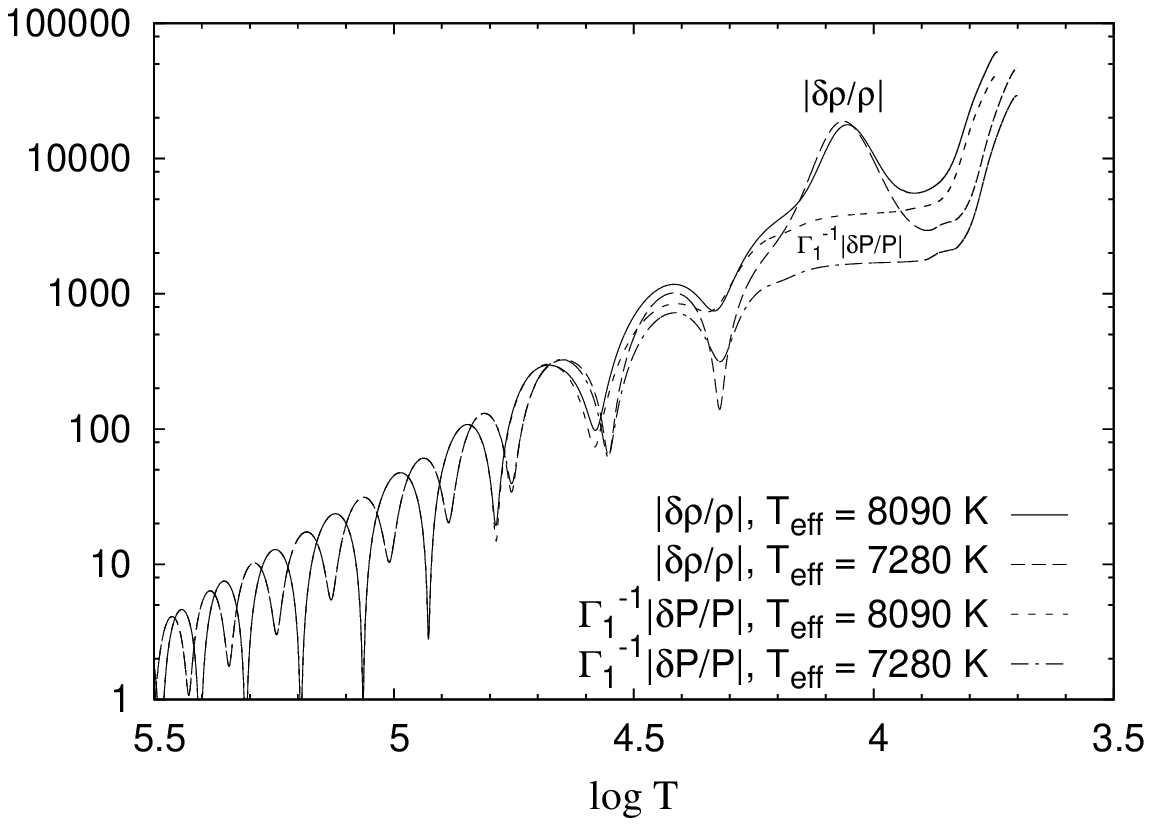}}
\caption{\label{drho-dp}$|\delta\rho/\rho|$ and $\Gamma_1^{-1}|\delta P/P|$ for the radial mode p$_{32}$ in two 1.6 M$_\odot$ models with different $T_{\rm eff}$.}
\end{figure}

The physical processes determining the instability strip red edge differ in our roAp
models compared with other $\kappa$-driven oscillators, such as $\delta$ Scuti stars. In the latter case,
the damping of the modes originates in the time-dependent coherent coupling with convection
\citep{dupret05}. In our purely radiative roAp models, we also predict a red edge, and must search
for another explanation.

As can be seen in Fig. \ref{kap}, the H$_{\rm I}$ opacity bump of cold models away from the instability strip is larger than it is
inside the instability strip. The large temperature gradient in this region implies that a density inversion exists that is
stronger for the cold model (Fig. \ref{rho}). {\it With this in mind, we could have expected more driving in the cold
model compared with the hot model (at least in a classical $\kappa$-mechanism vision of the problem) but
the converse instead occurs}. We also note that radiative damping is almost identical for the two models
(see Fig. \ref{work-red}),
because the temperature gradient and the eigenfunctions are similar in this region for different models.
The explanation of the red edge comes from a closer study of the eigenfunctions and their respective
phases in the H$_{\rm I}$ opacity bump.

As already explained above and shown in Figs. \ref{ds-dt} and \ref{drho-dp},
$|\delta s/c_p| \simeq |\delta T/T| \approx |\delta \rho/\rho| >> |\delta P/P|$ in the H$_{\rm I}$ opacity bump
region.
We also note in these figures, that the values obtained for $|\delta s/c_p|$ in this region
are close for the hot and
cold models. Figure \ref{drho-dp} illustrates however that the values of $|\delta P/P|$ are much smaller in the cold model than the hot model.
As mentioned before, it is more appropriate to consider the work expression in terms of $\delta P$ instead of $\delta\rho$ to be able to observe how much driving occurs in this region.

Since $|\delta P/P|$ is far smaller in the cold model, $|\Im(T\delta s^*\delta P/P)|$ is smaller and less work is performed
(see Eq.~\ref{workeq}). This smaller amount of driving cannot counterbalance the radiative
damping of deeper layers and all modes are stable.

The eigenfunctions have a complex non-adiabatic behavior in the superficial layers, and the interpretation that we propose of the behavior of $\delta P/P$ must be considered with caution. A possible origin of 
the differences found for $\delta P/P$ between the two models could be the densities (Fig.~\ref{rho}). 
According to the asymptotic theory, the following behavior of $P'/P$ is expected:
\begin{equation}
\label{asymptoteq}
\frac{P'}{P}(r)\:\propto\: \frac{(\rho c)^{1/2}}{P}\: F\left({\textstyle\int} \sigma/c\,dr\right)
\:\propto\: \frac{1}{\rho^{1/2}\,T^{3/4}}\: F\left({\textstyle\int} \sigma/c\,dr\right)\:,
\end{equation}
where $F$ is a cosine in propagation regions and an exponential in evanescent regions, $c$ is the sound speed, 
and $\sigma$ the angular frequency.
The asymptotic theory does not hold of course in the highly non-adiabatic superficial layers,
but $P'/P$ is an eigenfunction less affected by non-adiabatic effects because of the
control by the movement equation. At a given local temperature, Fig. \ref{rho} shows that the cold model has a larger 
density inversion than the hot model, because of its larger temperature gradient. 
Hence, we understand from Eq. (\ref{asymptoteq}) that $P'/P$ (and therefore $\delta P/P$) is smaller in the cold model. A second complementary explanation could also be the following. 
From Eq.~ (\ref{asymptoteq}), we see that,
in the asymptotic adiabatic regime, the scale height of the variation in the eigenfunctions is given by
$c/\sigma$, which is acceptable for $\delta P/P$. However, the scale height of variation in the other thermodynamic eigenfunctions is
$-{\rm d}r/{\rm d}\ln T$ because they are affected by the opacity bump. 
Since the cold model has a higher opacity (Fig. \ref{kap}) and therefore larger temperature gradient, the increase in 
$\delta P/P$ of\  
$\Delta\ln|\delta P/P|\approx (\sigma/c)\, \Delta r \approx (\sigma/c)\,({\rm d} r/{\rm d}\ln T)\, \Delta \ln T$
is expected to be smaller.
Hence, $\delta P/P$ cannot reach sufficiently high values around the H$_{\rm I}$ opacity bump of the cold model, 
and high order p-modes are stable.

\subsubsection{Shape and nodes of the eigenfunctions}
\label{nodesect}

\begin{figure}[h]
\resizebox{\hsize}{!}{\includegraphics{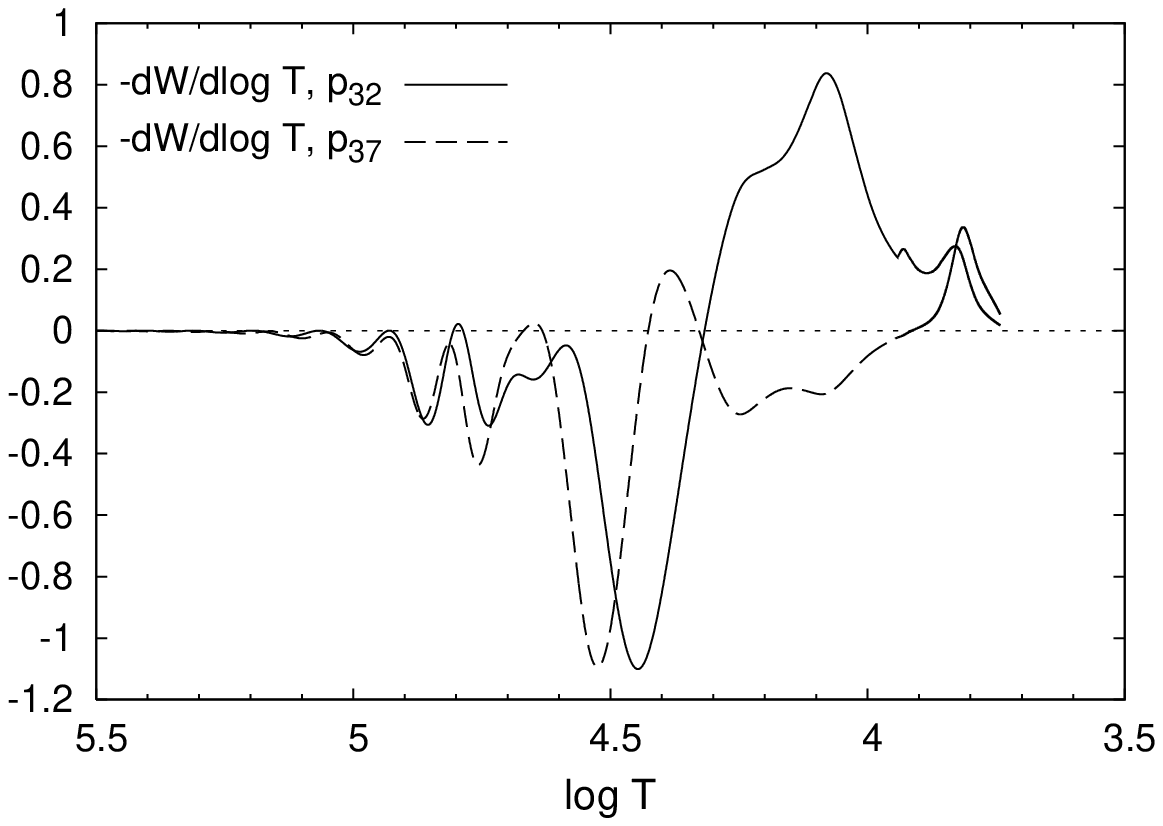}}
\caption{\label{dW-node}Comparison of the differential work $-{\rm d}W/{\rm d}\log T$ for the radial modes p$_{32}$
and p$_{37}$,  1.6 M$_\odot$ model with $T_{\rm eff}=8090$~K.}
\resizebox{\hsize}{!}{\includegraphics{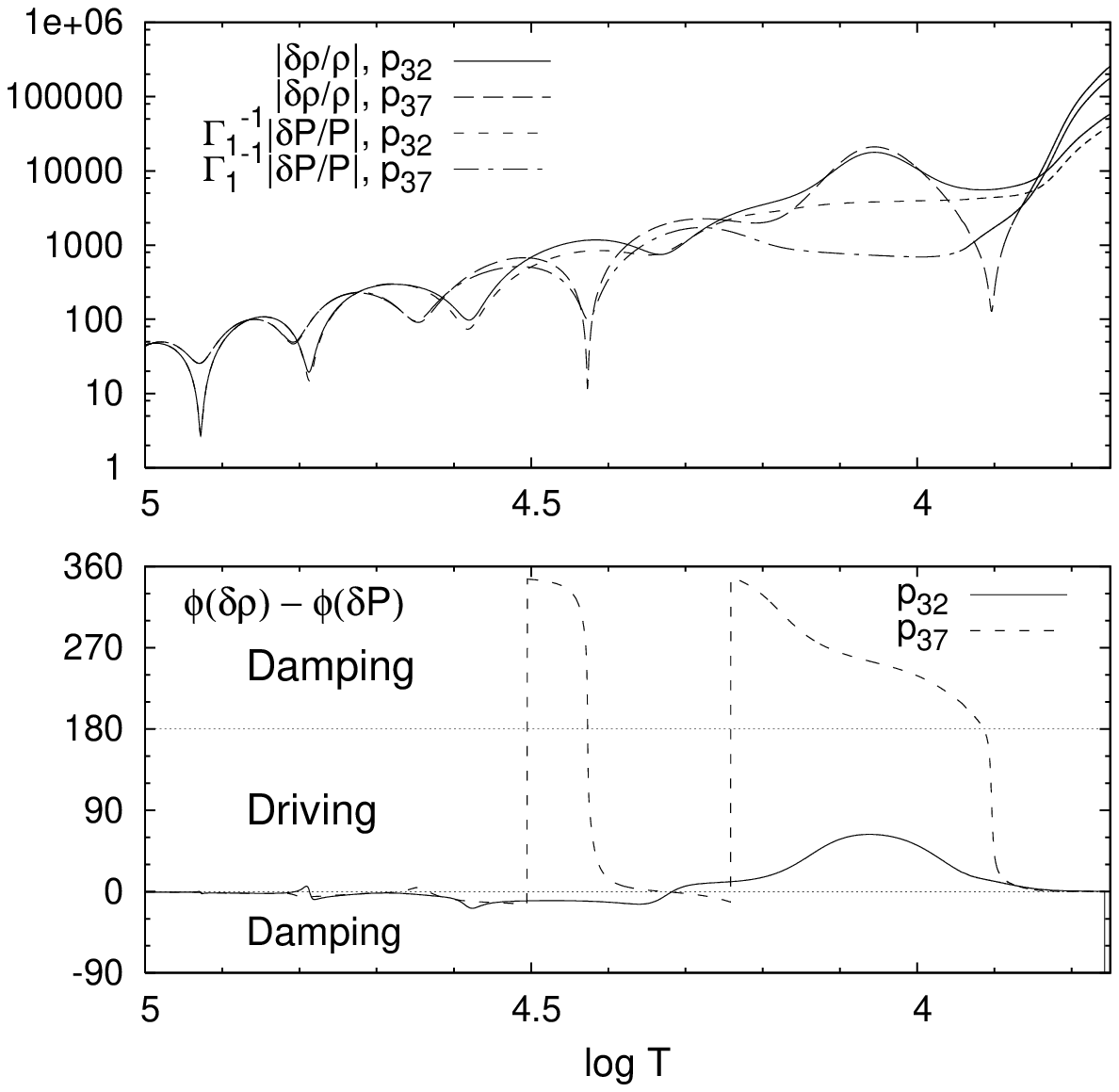}}
\caption{\label{drhoP-node}
$|\delta\rho/\rho|$ and $\Gamma_1^{-1}|\delta P/P|$ (top panel) and
$\phi(\delta\rho)-\phi(\delta P)$ (bottom panel) for the modes $p_{32}$ and $p_{37}$,  
1.6 M$_\odot$ model with $T_{\rm eff}=8090$~K.}
\end{figure}

We emphasize that the predictions concerning the driving and damping mechanisms in roAp stars are
extremely sensitive to the shape of the eigenfunctions. For these high radial order p-modes, nodes can
appear in both the driving region and the atmosphere. The latter phenomenon is not purely theoretical, since
it is observed in the line-profile variations of some roAp stars. 
As an example, we compare in Figs. \ref{dW-node} and \ref{drhoP-node} the results obtained 
for the same hot model as before, but for two
different modes: p$_{32}$ and p$_{37}$. We can see in Fig. \ref{dW-node} which presents $-{\rm d}W/{\rm d}\log T$, 
that the mode p$_{37}$ is damped almost everywhere ! To understand this result, 
in the top panel of Fig. \ref{drhoP-node}, we provide
$|\delta\rho/\rho|$ and $\Gamma_1^{-1}|\delta P/P|$,
and in the bottom panel $\phi(\delta\rho)-\phi(\delta P)$ (degrees) for the modes p$_{32}$ and p$_{37}$.
We recall (Eq. (\ref{workeq})) that regions in which $0^\circ<\phi(\delta \rho)-\phi(\delta P)<180^\circ$
drive the oscillations, and regions in which $180^\circ<\phi(\delta \rho)-\phi(\delta P)<360^\circ$
dampen the oscillations. The difference between the works shown in  Fig. \ref{dW-node}
is the presence of a node of $\delta\rho$ at $\log T\simeq 3.9$ for p$_{37}$, but not for p$_{32}$.
In both cases,  $\phi(\delta\rho)-\phi(\delta P)\simeq 0$ at the surface, because of the small
values of $\delta T/T$ due to the thermal boundary condition. However due to the location of the node of $\delta\rho$,
$\phi(\delta\rho)-\phi(\delta P)$ enters in the damping domain for $4.3>\log T> 3.9$, and
the mode p$_{37}$ is therefore stable.
The location of the nodes generates a type of windowing in the driving mechanism of roAp stars. The nodes must be at just the right place to allow the driving to occur. This windowing explains why the theoretical frequency
interval of unstable modes is small for any given roAp model (Fig.~\ref{freq}).

\subsection{Global metallicity influence}
\label{global}
\subsubsection{Instability strips}

\begin{figure*}
   \centering
   \includegraphics[width=0.42\textwidth]{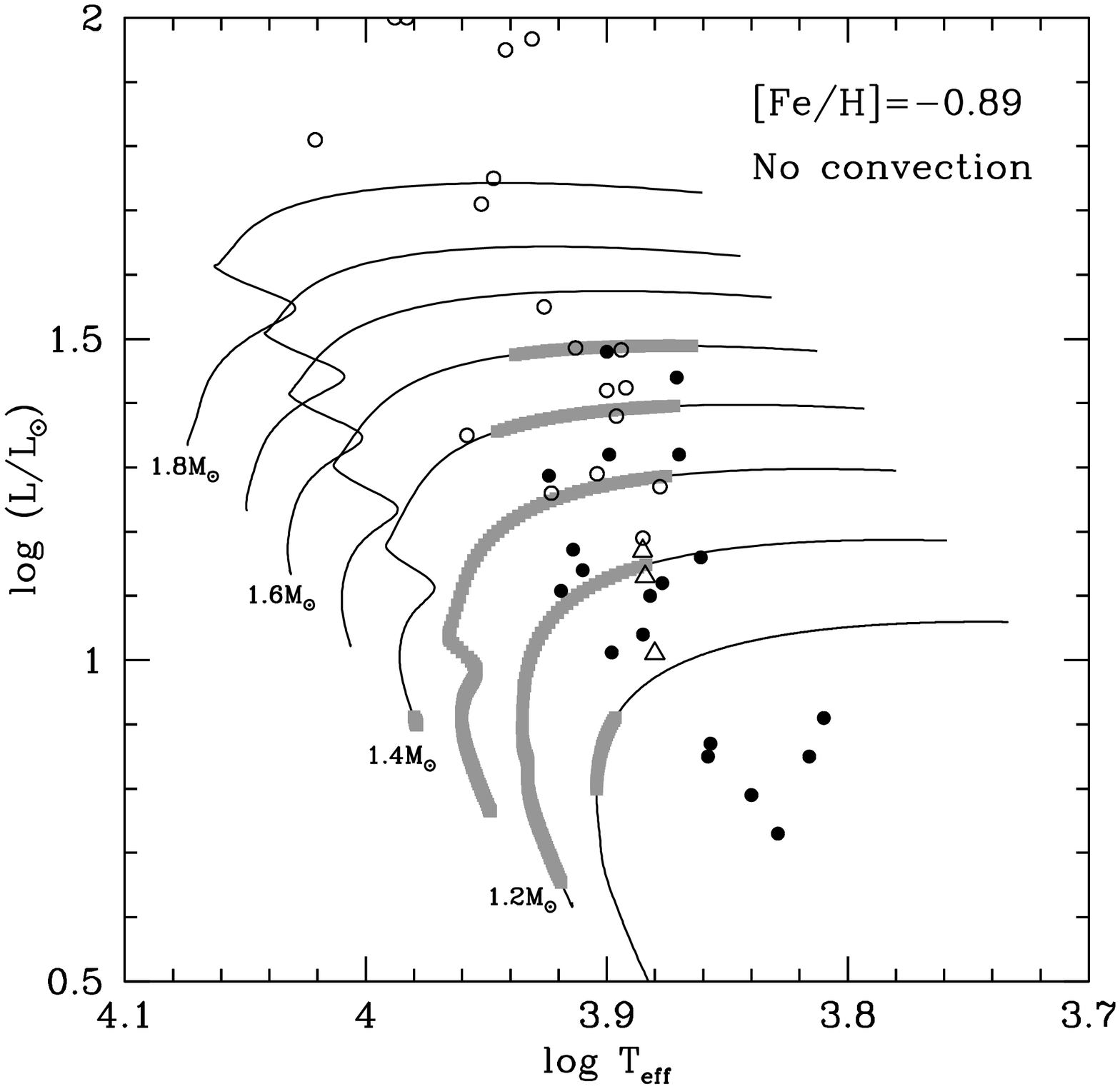}
   \includegraphics[width=0.42\textwidth]{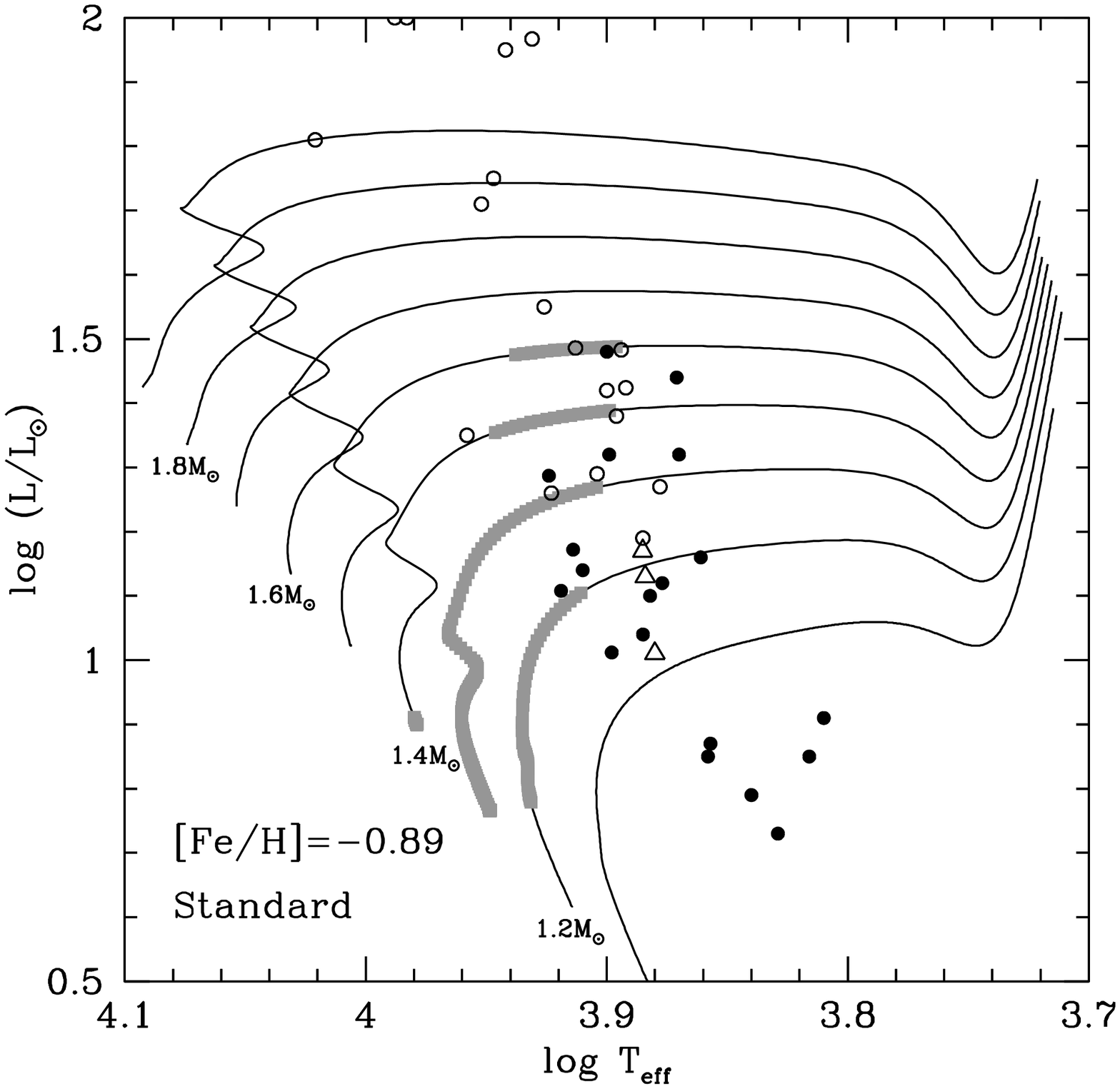}
   \includegraphics[width=0.42\textwidth]{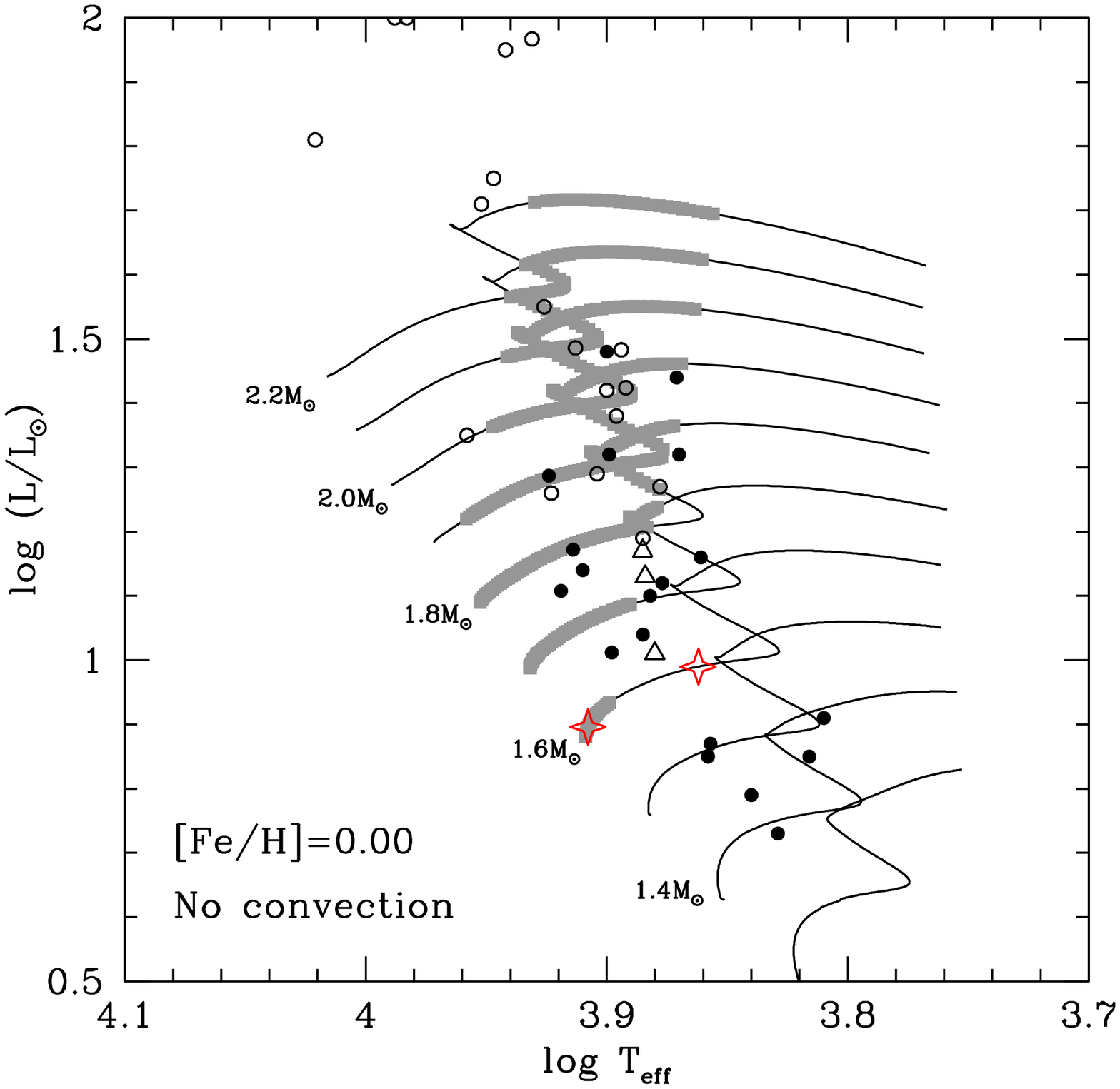}
   \includegraphics[width=0.42\textwidth]{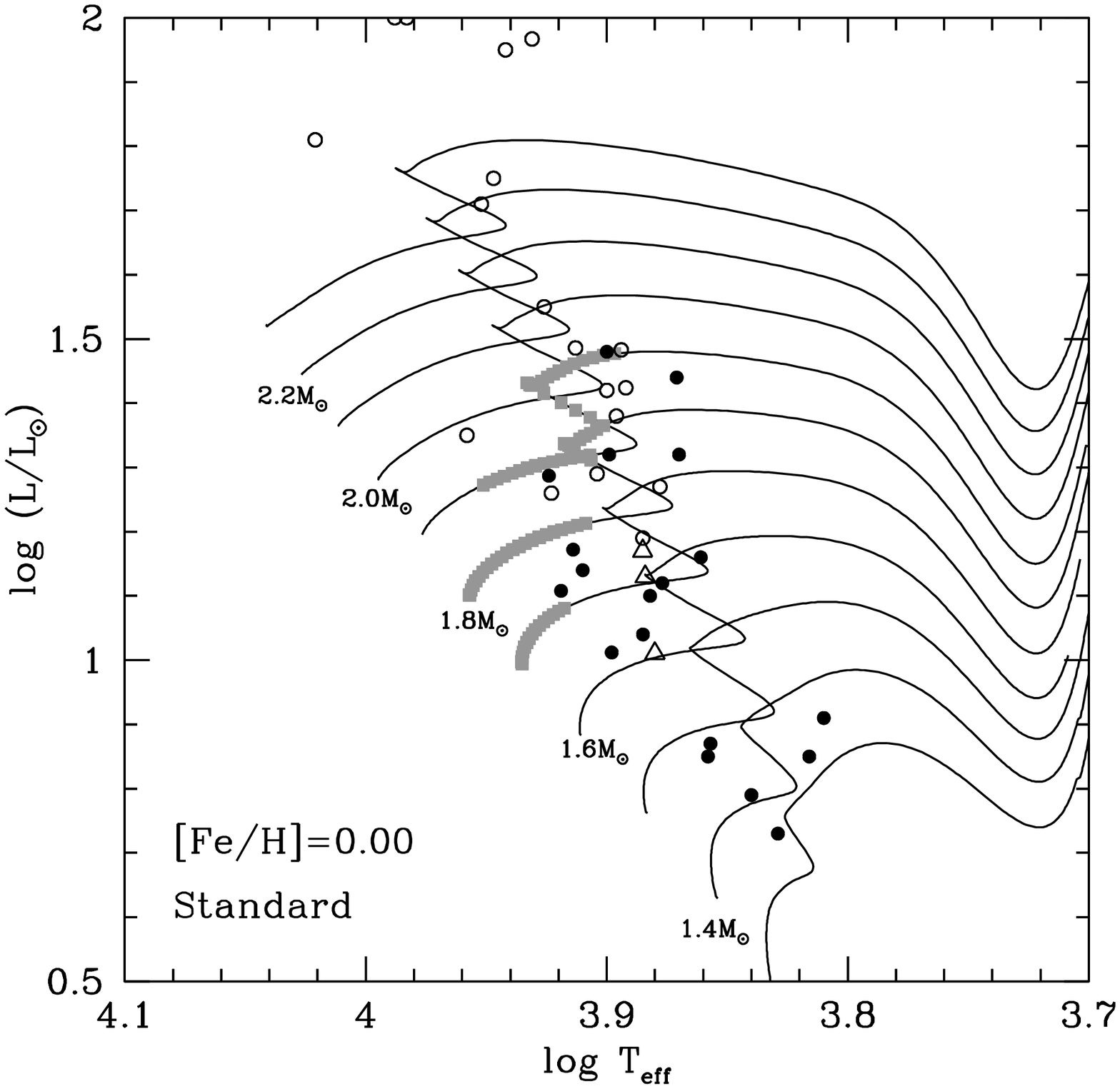}
   \includegraphics[width=0.42\textwidth]{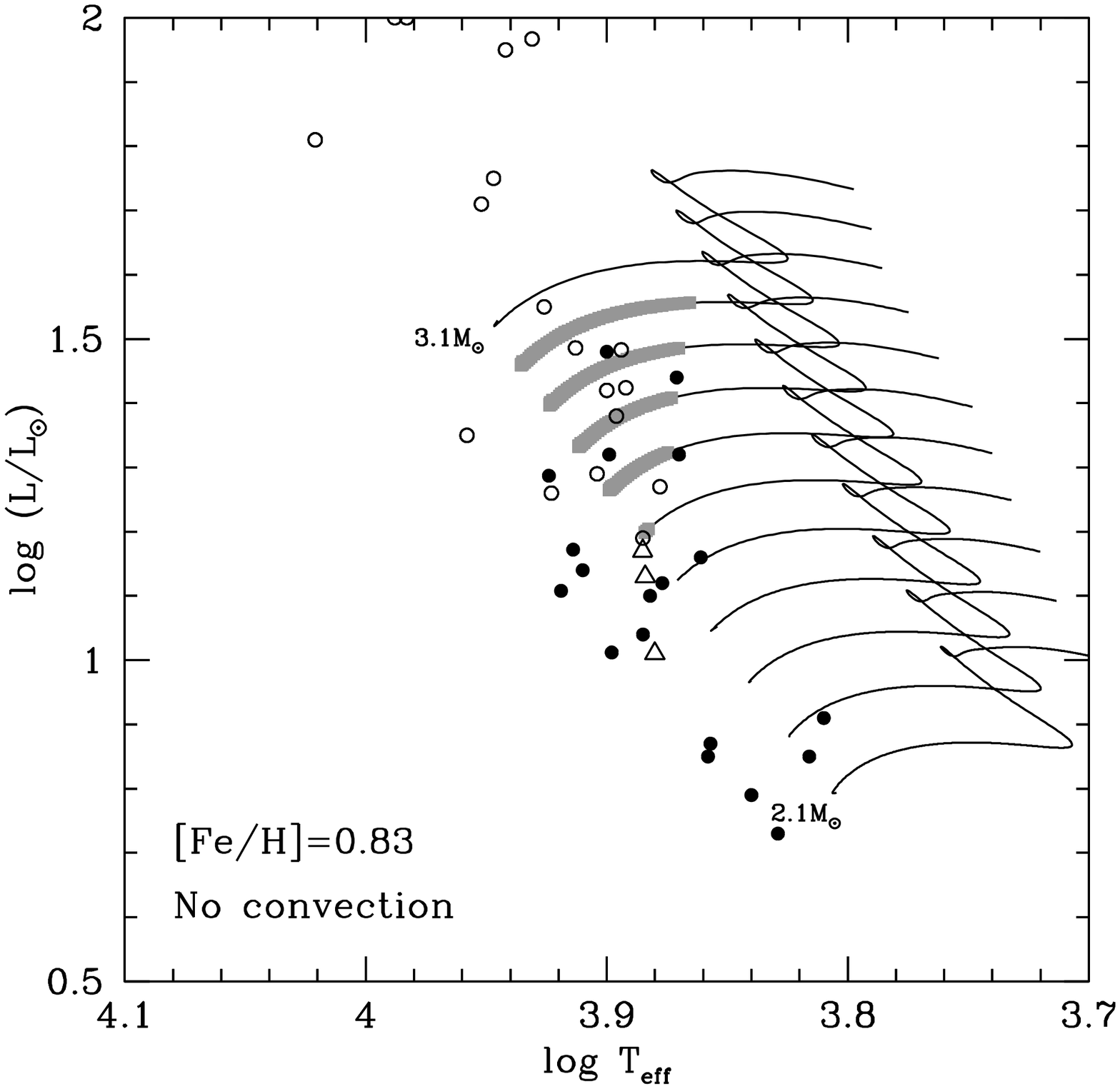}
   \includegraphics[width=0.42\textwidth]{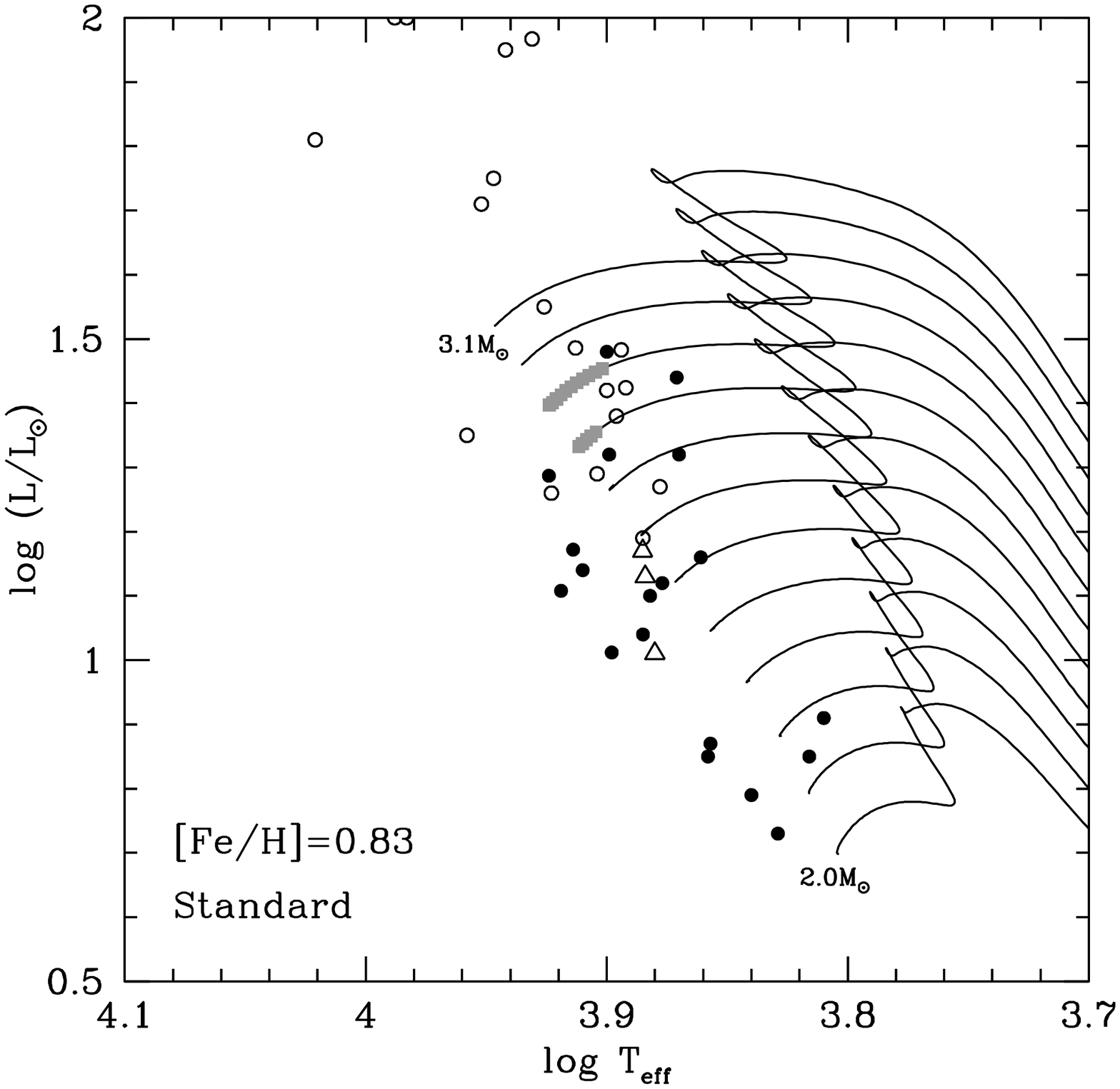}
   \caption{Evolutionary tracks and theoretical instability strips of model grids with different metallicities. Left column : models with convection suppressed in the envelope; right column : standard models (with convection). Upper panel : low metallicity models, middle panel : solar metallicity models, lower panel : metal-rich models. The full gray squares represent the models for which roAp-type modes are found to be excited. The circles and triangles represent the observations in Fig. \ref{diaghr}. For clarity, masses corresponding to the evolutionary tracks are given only for a few models. Each model grid is computed with a 0.1M${_\odot}$ step in mass. The two stars on the left middle figure correspond to the 1.6M$_{\odot}$ models compared in Fig. \ref{kap} to \ref{drhoP-node}.}
              \label{isz}%
   \end{figure*}

To test the influence of the global metallicity on the excitation mechanism of oscillations, we computed grids of models for three different metallicities: [Fe/H]= -0.89, 0.00 and 0.83 (the initial mass fraction of hydrogen is the same as previously X=0.71). Figure \ref{isz} shows the evolutionary tracks of the computed models. The left column shows the evolutionary tracks of models with convection suppressed and for comparison, the right column displays the results obtained for standard models (i.e. with convection). The full gray squares represent the models for which unstable roAp-type modes are found.

\begin{figure}
\resizebox{\hsize}{!}{\includegraphics{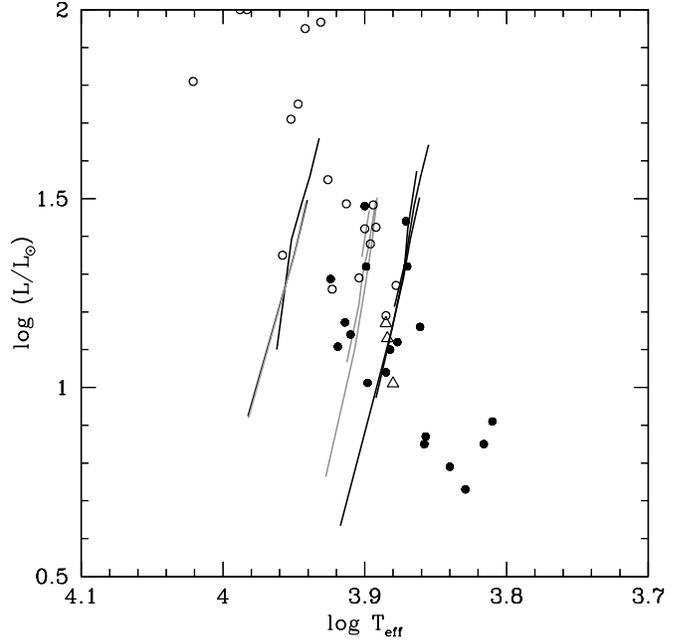}}
   \caption{Superposition of the instability strips deduced from models of Fig. \ref{isz}. Black lines are the instability strips derived from models grids with convection suppressed, gray lines the instability strips derived from standard models.}
              \label{isrecap}%
   \end{figure}

Figure \ref{isrecap} compares the position of the resulting instability strips. The black lines show the position of the instability strips derived from models with convection suppressed, and the gray lines the instability strips deduced from standard models. The figure illustrates how the suppression of convection increases the width of the theoretical instability strip towards cooler temperatures. In contrast, the global metallicity does not influence the position of the instability strip significantly. The models with solar metallicity, however, provide a closer agreement between observational points and theoretical instability strip (see Fig. \ref{isz}). The excitation of modes in stars with effective temperatures lower than 7400K ($\rm \log T_{eff} \le 3.87K$) is, however, not explained by these models.

\subsubsection{Comparison between calibrated models}
\label{compcal}

\begin{figure}
\resizebox{\hsize}{!}{\includegraphics{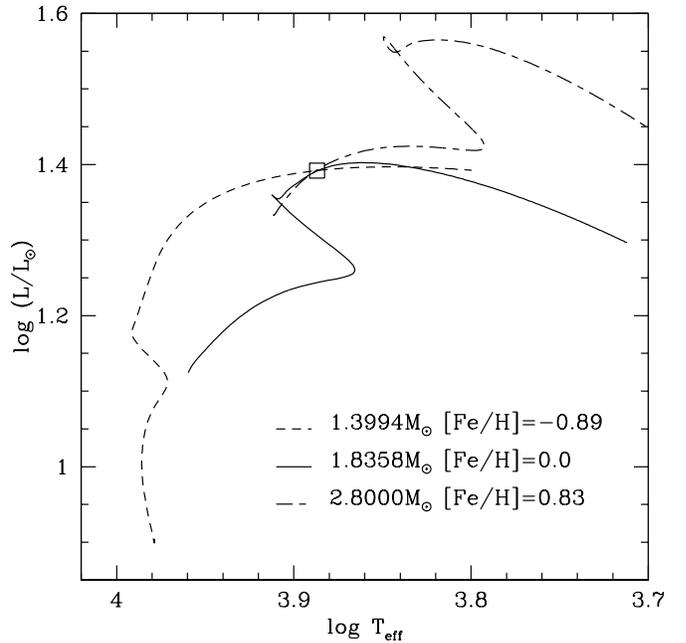}}
   \caption{Evolutionary tracks for three models with different masses and metallicities showing a common intersection (located by the open black square) in the HR diagram.}
              \label{diaghrg1}%
   \end{figure}

To study in detail the influence of the global metallicity on the excitation mechanism of pulsations, we compare three models of different masses and metallicities but located at similar position in the HR diagram. Figure \ref{diaghrg1} shows the evolutionary tracks of the three computed sequences. In the following, we compare the structures of the models located at the intersection of the evolutionary tracks. The characteristics of the three models are presented in Table \ref{table1}.

\begin{table}
\caption{Global characteristics of three models with different global metallicities and the corresponding frequency range of excited modes.}
\medskip
\centering
\begin{tabular}{ccccc}
\hline\noalign{\smallskip}
M (M$_{\odot}$)& [Fe/H] &  T$_{\rm eff}$ (K) & $\rm \log (L/L_{\odot})$ &Excitation range\\
\hline\noalign{\smallskip}
1.3994  & -0.89 & 7715 & 1.3919 & 0.666-0.698 mHz \\
1.8358 & 0.00 & 7690 & 1.3925 & 0.902-1.009 mHz\\
2.8000 & 0.83 & 7714 & 1.3918 & 1.526-1.569 mHz\\
\noalign{\smallskip}
\hline
\end{tabular}
\label{table1}
\end{table}

\begin{figure}
\resizebox{\hsize}{!}{\includegraphics{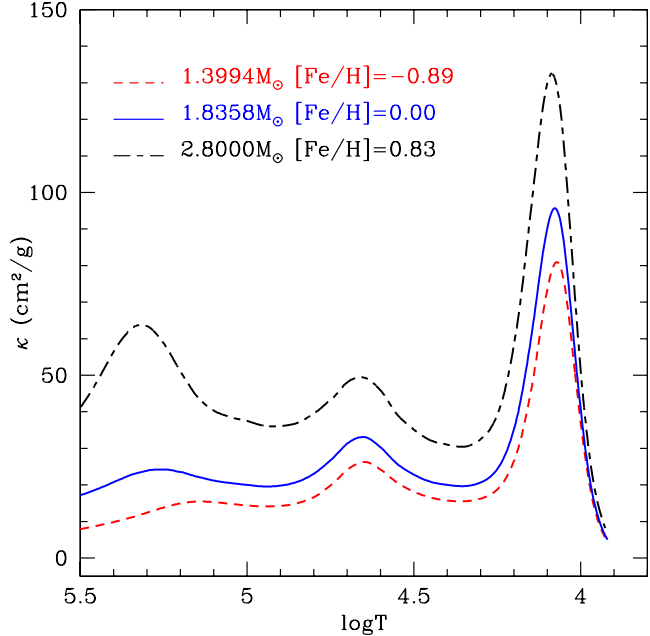}}
   \caption{Opacity profiles in models with different masses and metallicities but located at the same point in the HR diagram.}
              \label{kappag1}%
   \end{figure}

Figure \ref{kappag1} compares the opacity distribution in the three models. The global metallicity strongly affects the opacity profiles. In particular, it affects the opacity close to $\rm \log T \simeq 4.1$, i.e. in the H ionization zone. In the context of a ``normal'' $\kappa$-mechanism driving, we should expect the metallicity to affect the excitation of the modes.

\begin{figure}[h]
\resizebox{\hsize}{!}{\includegraphics{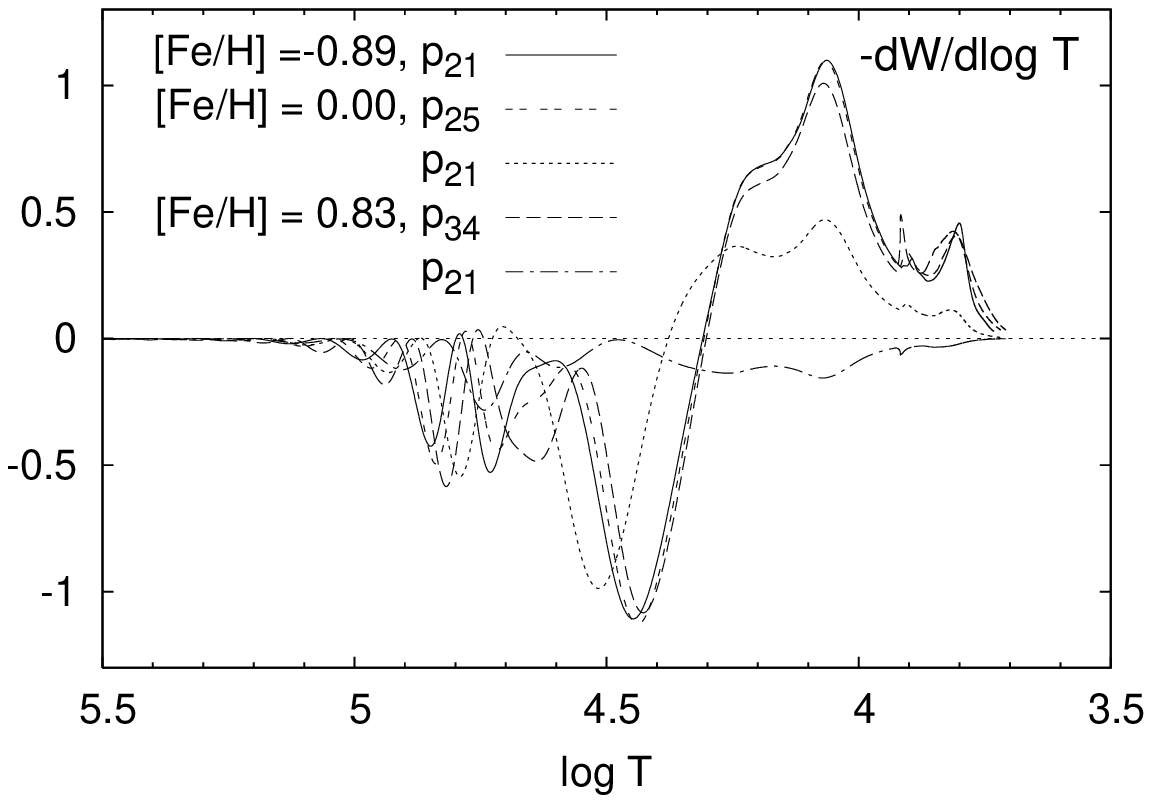}}
\caption{\label{dW-cal}Comparison of the differential work $-{\rm d}W/{\rm d}\log T$ for three models with different global metallicities ([Fe/H]=-0.89, [Fe/H]=0.00 and [Fe/H]=0.83) at the same location in the HR diagram.}
\end{figure}

However, this does not appear to be the case, since the three models have excited high
order p-modes. To understand more clearly what is happening, we provide in
Fig. \ref{dW-cal} the differential work obtained for these three models.
The modes considered in this figure are:
\begin{itemize}
\item for the [Fe/H]=-0.89 model: p$_{21}$ ($\nu=0.698$ mHz, excited),
\item for the [Fe/H]=0.00 model: p$_{25}$ ($\nu=0.938$ mHz, excited)
and p$_{21}$ ($\nu=0.792$ mHz, stable),
\item for the [Fe/H]=0.83 model: p$_{34}$ ($\nu=1.569$ mHz, excited)
and p$_{21}$ ($\nu=0.979$ mHz, stable).
\end{itemize}

\begin{figure}[h]
\resizebox{\hsize}{!}{\includegraphics{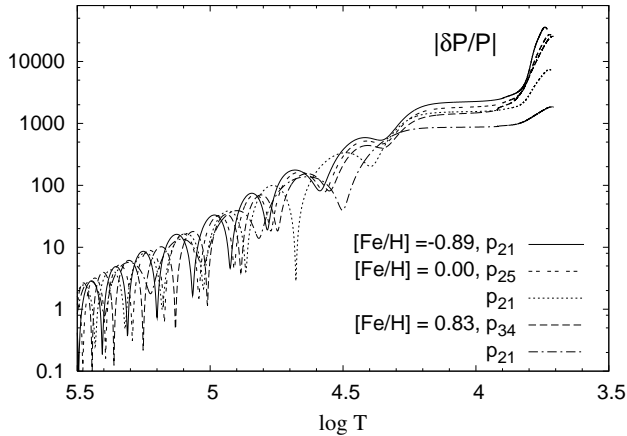}}
\caption{\label{dp-cal}Comparison of $|\delta P/P|$ for three models with different global metallicities ([Fe/H]=-0.89, [Fe/H]=0.00 and [Fe/H]=0.83) at the same location in the HR diagram.}
\end{figure}

We can see in Fig. \ref{dW-cal} that the driving is similar for the three excited modes, although they have significantly different radial orders. On the other hand, no
significant driving occurs for the p$_{21}$ mode of the [Fe/H]=0.00 and [Fe/H]=0.83 models. As we explained in Sect.~\ref{nodesect}, a key point for the driving
of roAp modes is the last node location of the eigenfunction
$|\delta P/P|$. In Fig. \ref{dp-cal}, we show $|\delta P/P|$ for the same modes
and models as before. We can see that the last node of $|\delta P/P|$ is
located at the same temperature for all three excited modes. To have efficient
driving, it must be located at
$\log T_1\simeq 4.35$, just between the H$_{\rm I}$ and He$_{\rm II}$ opacity bumps.
The location of this last node corresponds to a specific value
of the acoustic depth given by
$\tau_c=\int \sigma/c \;{\rm d}r=\int_{T_1}^{T_2} \sigma/c \,({\rm d}r/{\rm d}T)\:{\rm d}T$, where T$_1$ is the temperature at the last node, and T$_2$ is the surface temperature. Conversely, from the relation:
\begin{equation}
\sigma=\tau_c\;/\int_{T_1}^{T_2} c^{-1} \,({\rm d}r/{\rm d}T)\:{\rm d}T\;.
\label{taueq}
\end{equation}
we can infer the frequency of the mode whose last node is located at a fixed temperature (T$_1$) and a fixed acoutic depth ($\tau_c$).

This simple relation allows us to understand how changes in the metallicity of the models
change the range of unstable modes. Increasing $Z$ increases the opacity significantly
as shown by Fig.~\ref{kappag1}, and therefore the temperature gradient. Hence, $\sigma$ must
increase, according to Eq.~(\ref{taueq}). In other words, as ${\rm d} T/{\rm d}r$
increases, the size of the opacity bump region decreases, so that we have to decrease the
wavelength by increasing the frequency to ensure that the last node remains at the same temperature.
In Fig.~\ref{dp-cal}, the location of the last node of the p$_{21}$ modes confirms that
this reasoning is correct. If the frequency of the mode is not increased sufficiently,
the size of the opacity bump region decreases more rapidly than the wavelength so that
the last node is located at higher temperature (at $\log T\simeq 4.4$, dotted curve for
the  [Fe/H]=0.00 model, and at $\log T\simeq 4.5$, dot-dashed curve for the [Fe/H]=0.83 model), which inhibits the driving.

Since the location of the last node is the critical point for the driving, and since it is always
possible to find a mode with this location, we can understand why the location
of the theoretical instability strip is poorly sensitive to changes in the structure model,
even if they strongly affect the opacity bump in the driving region.

\section{Conclusions}
Observations appear to imply that a relationship exists between the excitation mechanism of pulsations in roAp stars and their heavy element distribution (cf Sect. \ref{verticalstrat}). On the other hand, the region that plays a major role in driving the roAp high order p-modes corresponds to the last opacity bump around $\rm \log T_{eff} \simeq 4.1$. It has been assumed that this opacity bump was due mainly to the partial ionization of hydrogen. However, our study has demonstrated that the contribution of metals (typically iron) is significant. We have therefore determined how the theoretical instability strips of roAp stars is affected by a change in metallicity. Models with different global metallicity (Sect. \ref{global}), local enhancement of metals (Appendix \ref{localZ}, available in electronic form only), or iron (Appendix \ref{localFe}, available in electronic form only), were considered. The local enhancements are expected to be produced by the competing effects of gravitational settling and radiative levitation. Although the last opacity bump of these models is affected significantly by these changes, it appears that the theoretical instability strips are only poorly affected. Our models reproduce well the blue edge of the roAp instability strip, although the red edge deduced from the models is always too hot, so that our instability strips do not account for the cold roAp stars. 

We find that the driving mechanism of roAp high order p-modes differs qualitatively from the weakly nonadiabatic $\kappa$-mechanism acting in other classical pulsators. The shape of the eigenfunctions and, in particular, the location of their last node is critical. To allow efficient driving to occur, the last node of $\delta P/P$ must be located between the two last opacity bumps at $\rm \log T \simeq 4.3$. Changes in the structure of the superficial layers affect the driving less than the eigenfunctions. It follows that changing the opacity affects the driving differently than the classical $\kappa$-mechanism. In some cases, what we could refer to as an ``inverse $\kappa$-mechanism'' can occur: increasing the opacity and thus the temperature gradient leads to a density inversion close to the photosphere; the larger temperature gradient and density lead to smaller $\delta P/P$ and therefore a smaller amount of driving, which produces exactly the opposite effects of the classical $\kappa$-mechanism. This explains why the theoretical instability strips are less affected by the changes in our models than expected on the basis of the classical $\kappa$-mechanism. 

The weak effect of a drastic increase in opacity in the driving region is a surprising result, which led us to investigate the details in the $\kappa$-mechanism at work in such stars. In the future, we propose to consider additional changes to the models and observe how they affect the driving of the roAp high order p-modes. In particular, we intend to consider in more detail local changes in the He abundance of different regions (e.g. HeII and HeI ionization regions).

\begin{acknowledgements}
S.T. was supported by ESA-PRODEX ``CoRoT Preparation to exploitation I'' through the grants C90135.
J.F. acknowledges support from NSF grant 
AST-0239590 and support from Grant No. 
EIA-0216178 and Grant No. EPS-0236913 with matching support from the
State of Kansas and the Wichita State University 
High Performance Computing Center.

\end{acknowledgements}

\bibliographystyle{aa}
\bibliography{0494}

\Online

\begin{appendix}
\section{Local metallicity enhancement}
\label{localZ}
As described in Sect. \ref{verticalstrat}, there is observational evidence for vertical stratification in cool magnetic Ap stars, and for accumulation of some heavy elements in the lower atmosphere. Moreover, the coolest magnetic Ap stars have underabundances in some iron peak elements, which could also be an indication of accumulation of these elements somewhere below the surface.

Diffusion calculations in model atmospheres by \citet{leblanc04} suggested that, in cool Ap stars a high concentration of iron-peak elements occurs in low atmospheric layers and that it is plausible to expect a sharp decrease in Fe abundance above an optical depth $\log \tau_{5000} \simeq -1$.

\begin{figure*}[t]
   \centering
   \includegraphics[width=0.45\textwidth]{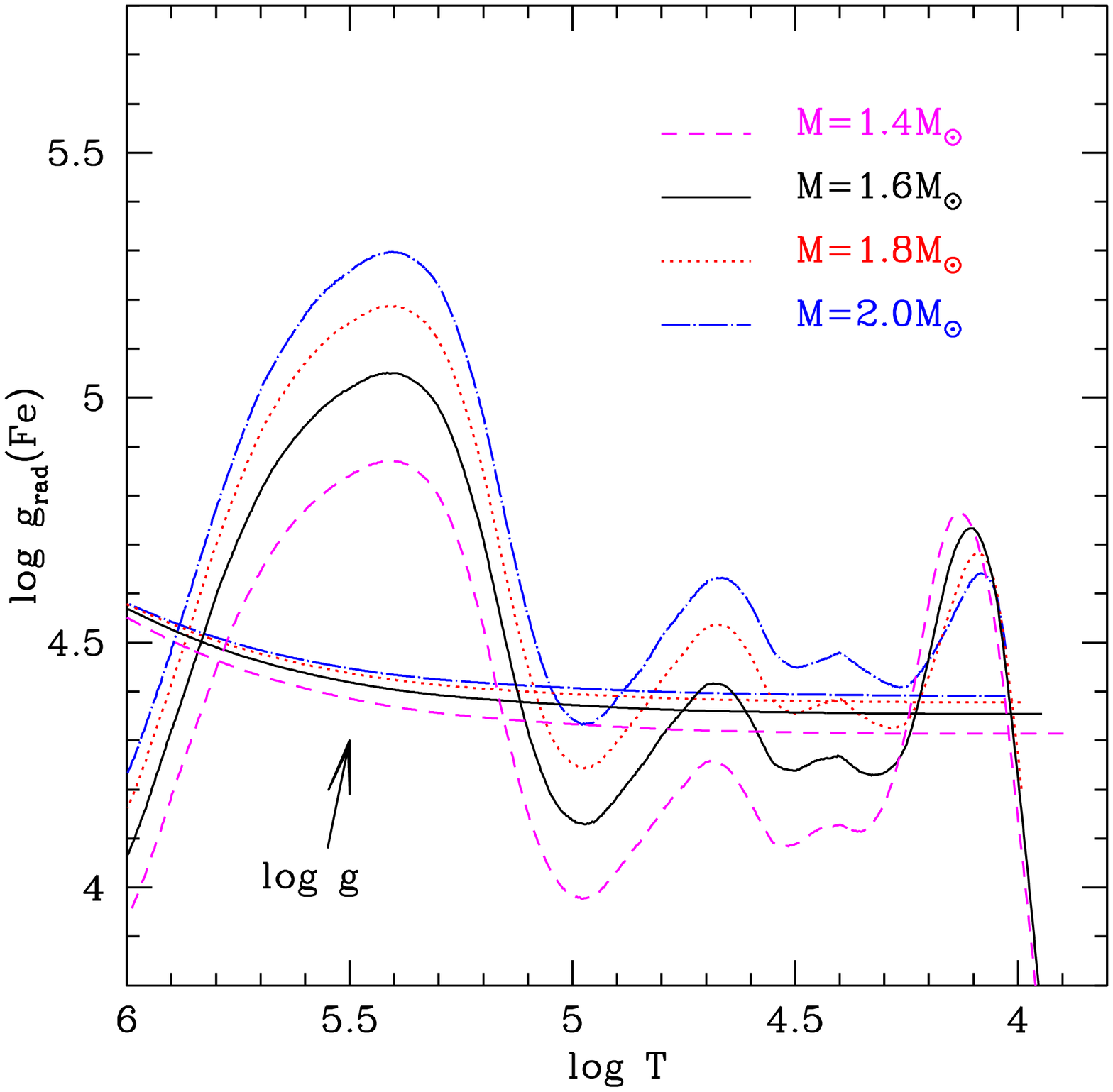}
   \includegraphics[width=0.45\textwidth]{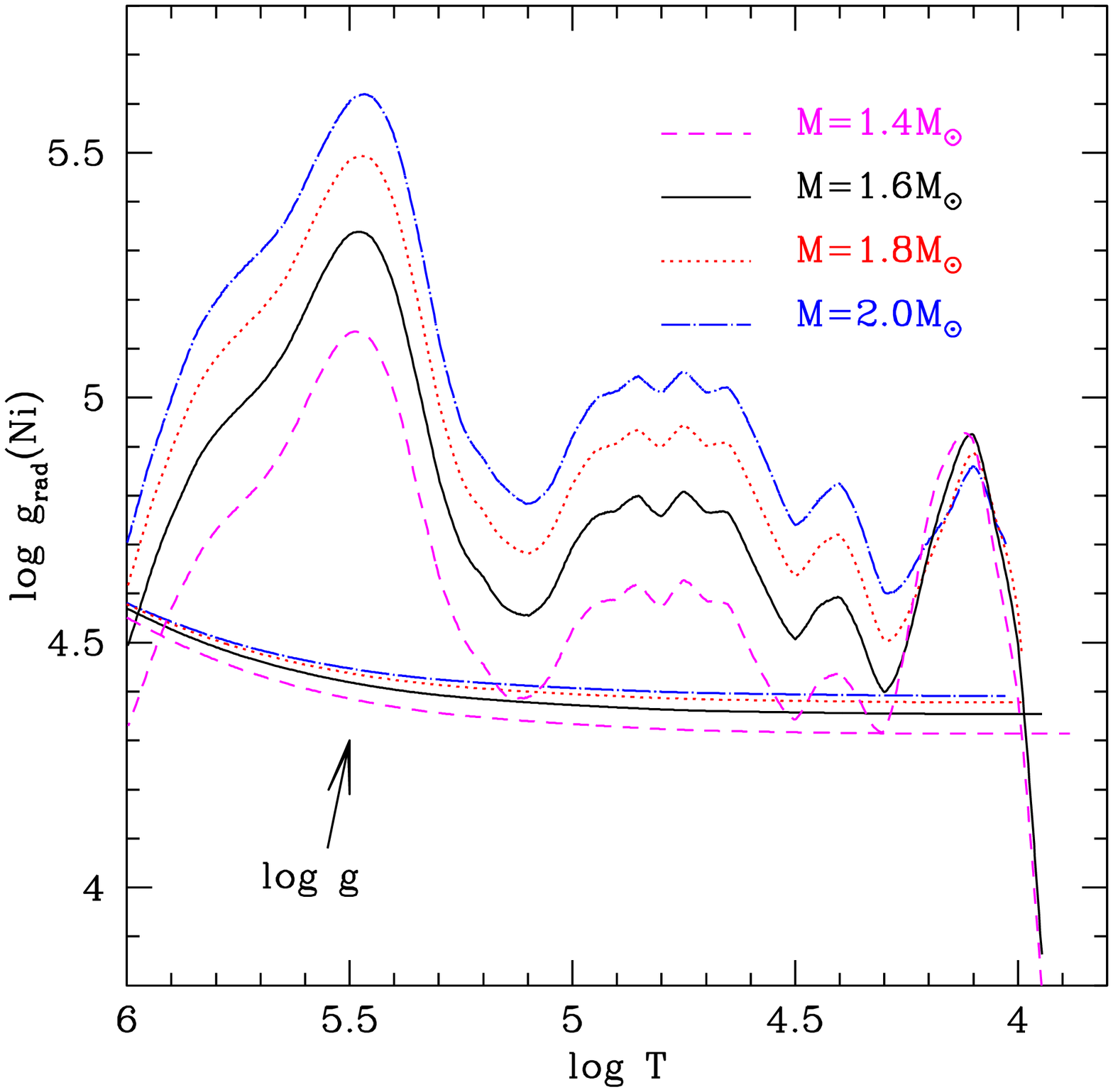}
   \includegraphics[width=0.45\textwidth]{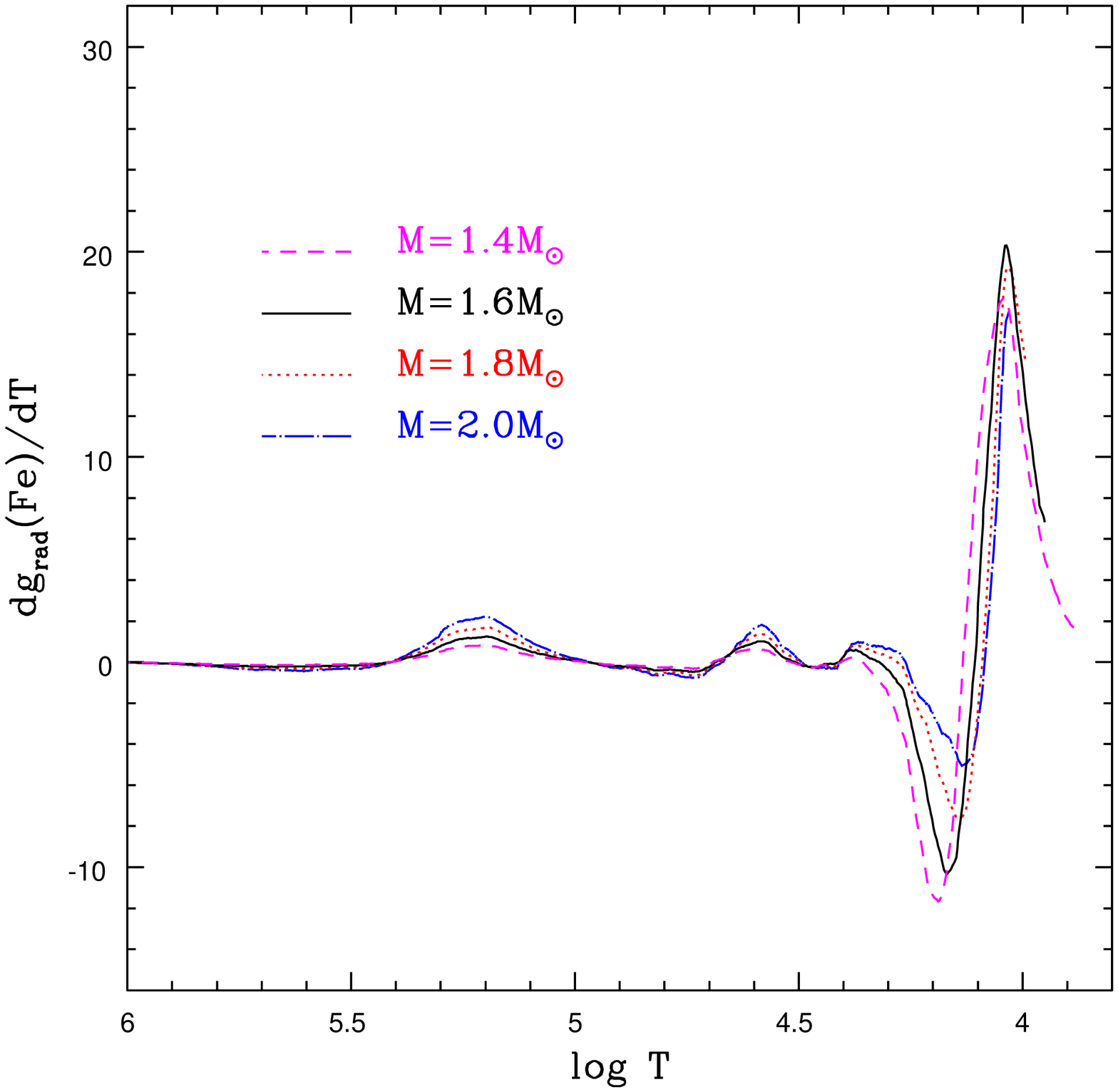}
   \includegraphics[width=0.45\textwidth]{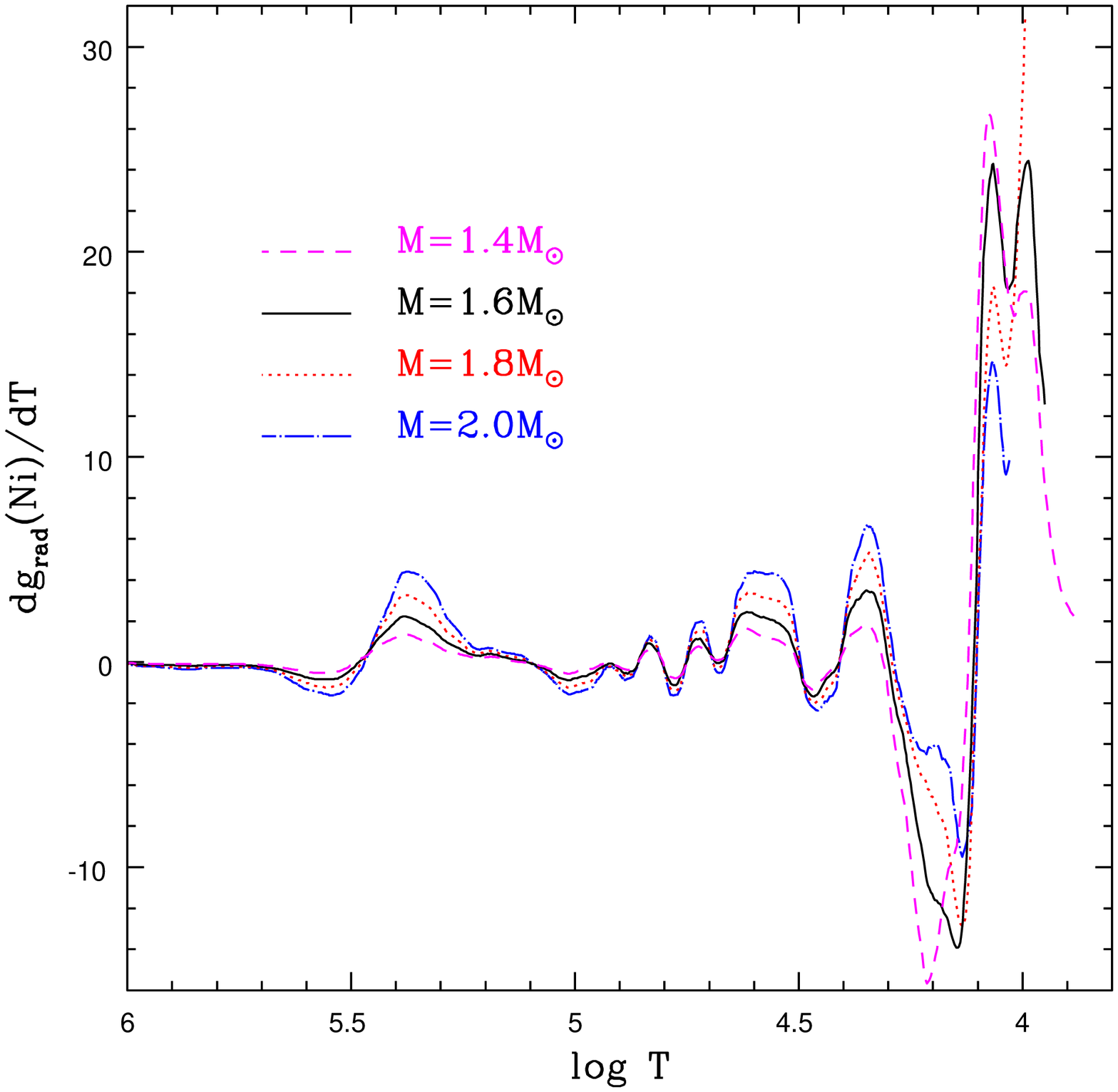}
   \caption{Upper panels : radiative accelerations on iron and nickel versus temperature inside four ZAMS models with different masses. The nearly horizontal curves represents the local gravity. Lower panels : derivatives of the radiative accelerations on iron and nickel. These models are computed with X=0.71 and [Fe/H]=0.00.}
              \label{grad}%
   \end{figure*}

The present version of Cl\'es does not allow us to compute models for A stars that include consistent computations of radiative acceleration effects. We therefore model the effects of microscopic diffusion in a parametric way.
In A stars, microscopic diffusion is expected to lead to accumulation of metals in some regions and to metal depletion in others. To locate these regions, we computed the radiative accelerations on Fe and Ni in several ZAMS models. The radiative accelerations were computed using the OPCD\_2.1 package taken from the website\footnotemark of the Opacity Project \citep[see][and references therein]{seaton05} \footnotetext{http://vizier.u-strasbg.fr/topbase/op.html}. The OPCD\_2.1 package contains data, codes and instructions for computing Rosseland mean opacities and radiative accelerations for any given stellar structure.
Using this package, we computed the radiative acceleration on iron and nickel in ZAMS models with masses between 1.4 and 2.0M$_{\odot}$. 
The initial hydrogen mass fraction of these models was X=0.71 and the metallicity was of the solar value. The upper panels of Fig. \ref{grad} compare the radiative accelerations on iron and nickel to the local gravity in 1.4, 1.6, 1.8, and 2.0 M$_{\odot}$ ZAMS models. The lower panels show the derivatives of the radiative accelerations. Iron and nickel are expected to accumulate in regions where the derivatives of the radiative accelerations are positive: we therefore expect iron to accumulate close to $\rm \log T=4.05$, 4.4, 4.6, and 5.2, and nickel to accumulate just below the surface (down to $\rm \log T=4.1$), close to $\rm \log T=4.35$, 4.6 and 5.4, at least during the early main-sequence phase. 

To study the influence of diffusion-induced iron/nickel accumulations on the excitation mechanism of magnetic Ap stars, we adopted an approach of applying the Fe and Ni enhancements to the entire mixture. We computed models in which parametric metal profiles, centered on chosen temperatures were introduced to simulate, qualitatively, iron peak element accumulations. Since the relevant regions for the driving of pulsations are the external layers where hydrogen and helium experience their ionization, we chose to study the influence of metal accumulation down to $\rm \log T=4.6$.

\begin{figure}
\resizebox{\hsize}{!}{\includegraphics[width=0.45\textwidth]{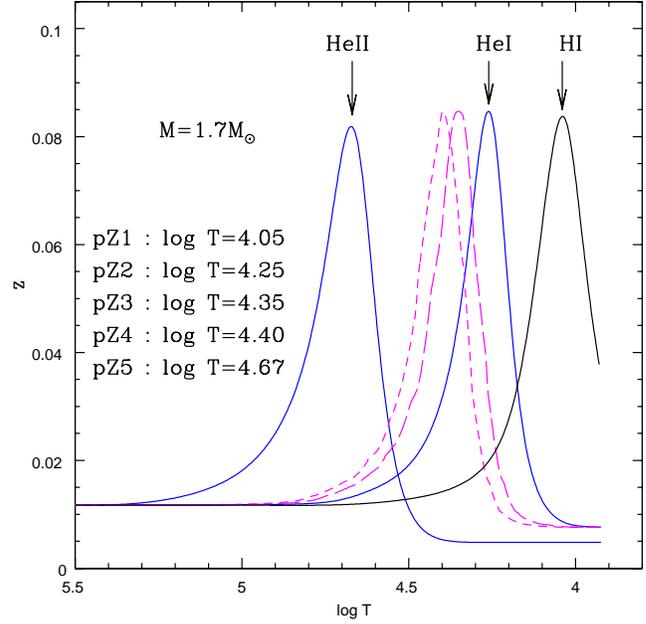}}
   \caption{Metal accumulation profiles (mass fraction) introduced in 1.7M$_\odot$ models. The model labelled pZ1 present a metal accumulation centered on the HI ionization region. In the pZ2 and pZ5 models the accumulation is centered respectively on the HeI and HeII accumulation region. The pZ3 and pZ4 models present an accumulation centered respectively on $\rm \log T=4.35$ and $\rm \log T=4.40$. }
              \label{profilZ}%
   \end{figure}

Our parametrization of the metal profiles was inspired by Eq. 4 of \citet{balmforth01}.  To center the metal accumulation profile on the HI ionization region, we used the following formula :
$$ Z=Z_s+A[4 x_{12}(1-x_{12})]^{\frac{1}{2}}+x_{12}(Z_i-Z_s)$$
where the profile centered on the HeI ionization region obeys the law :
$$ Z=Z_s+A[4 x_{21}(1-x_{21})]^{\frac{1}{2}}+(1-x_{21})(Z_i-Z_s)$$
and the profile centered on the HeII ionization region results from:
$$ Z=Z_s+A[4 x_{23}(1-x_{23})]^{\frac{1}{2}}+x_{23}(Z_i-Z_s)$$
where x$_{ij}$ is the ionization fraction of element i (i being its nuclear charge) in the (j-1)-th state of ionization. As an example, x$_{23}$ is the ionization fraction of doubly ionized helium. The parameters A and Z$_s$ are chosen arbitrarily to equal 0.075 and 0.65$\times$0.0117 respectively.
To obtain profiles centered on a given temperature (different from those of HI, HeI and HeII ionization regions), we use one of the aforementioned parametrization but we shift the profile towards the chosen temperature.
Figure \ref{profilZ} displays the profiles introduced into the various computed grids of models. The results are shown only for the 1.7M$_{\odot}$ models; the profiles introduced into the other masses originate in the same parametrizations and are similar.

\begin{figure*}[h]
   \centering
   \includegraphics[width=0.45\textwidth]{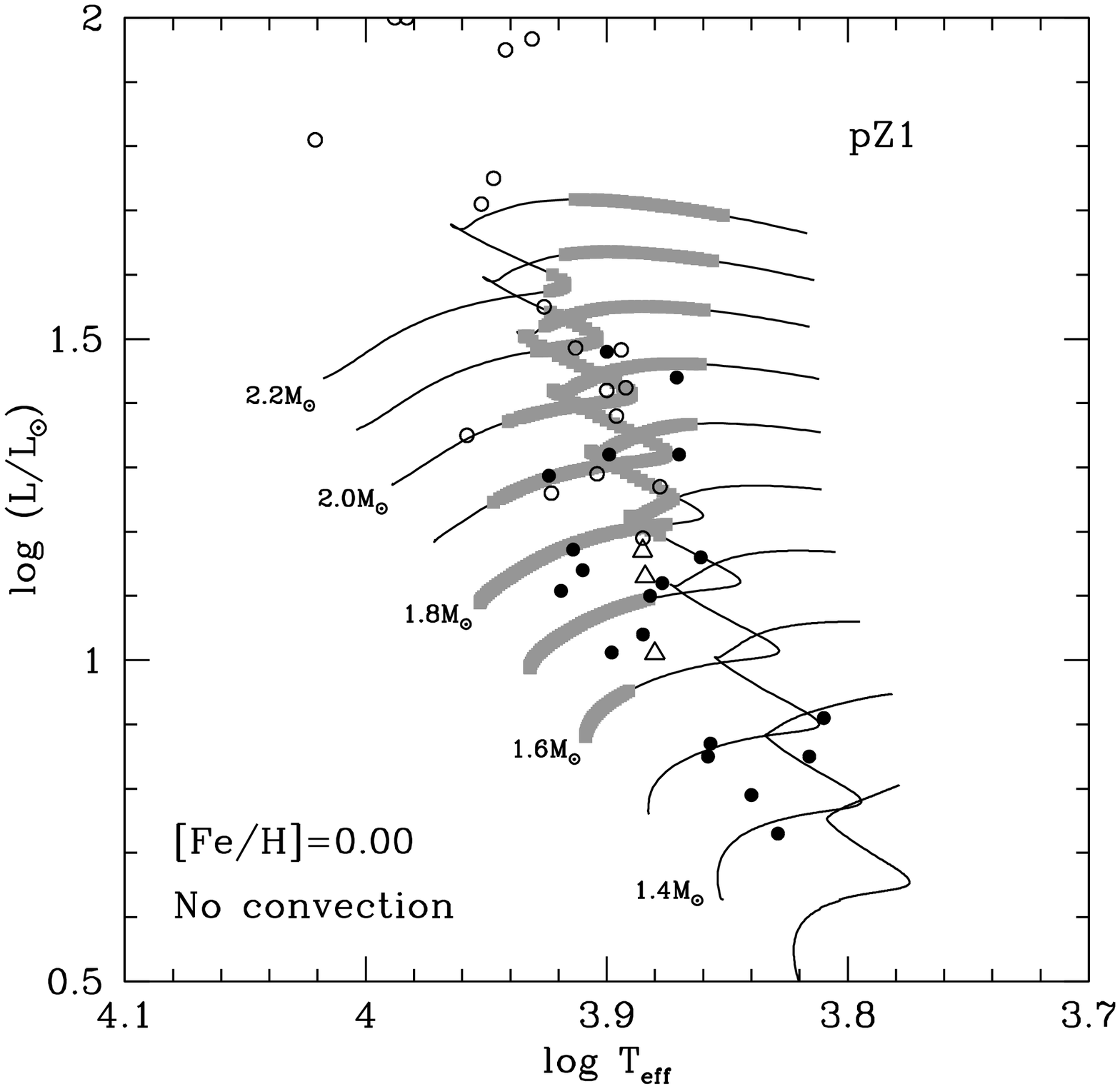}
   \includegraphics[width=0.45\textwidth]{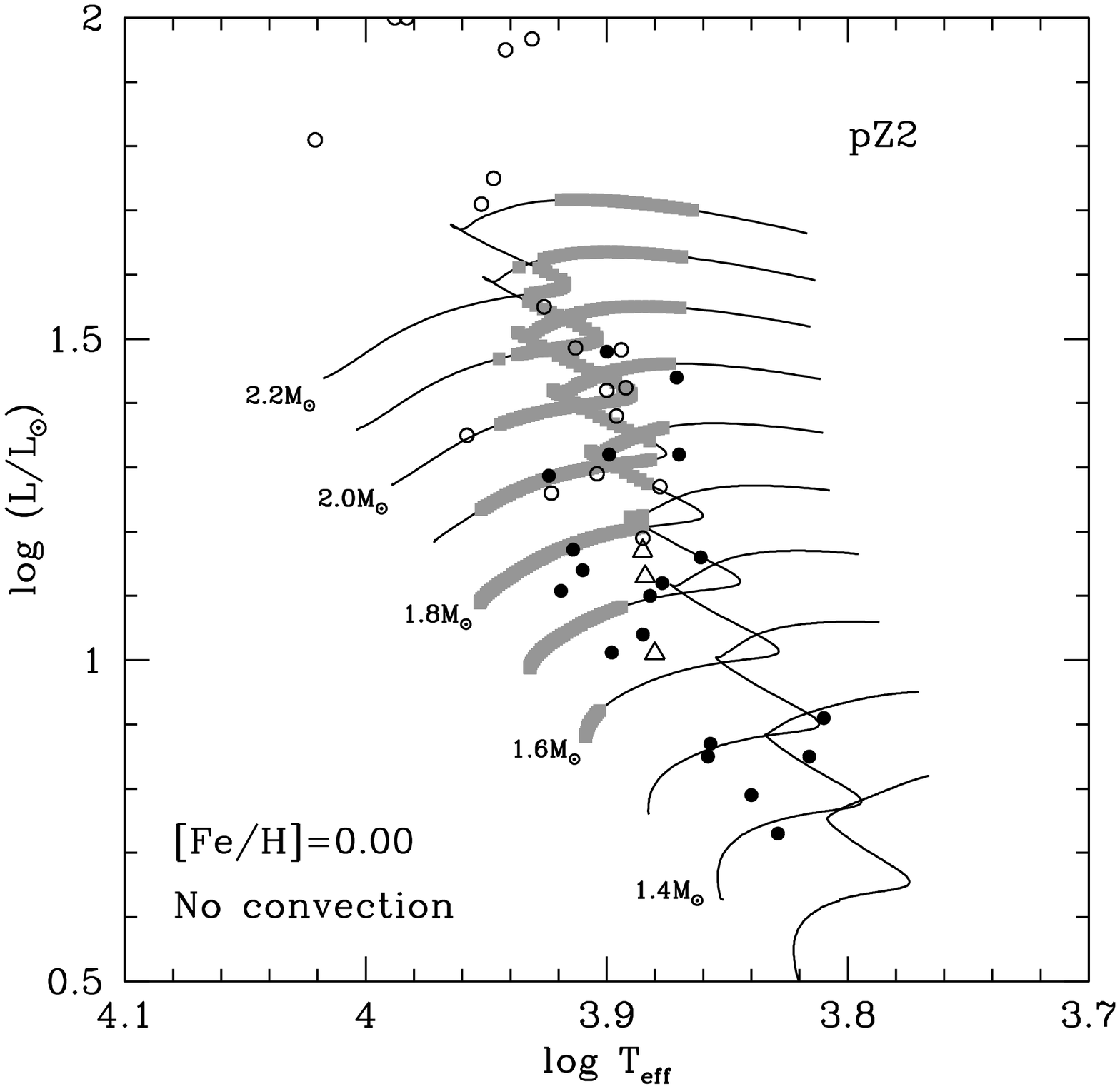}
   \includegraphics[width=0.45\textwidth]{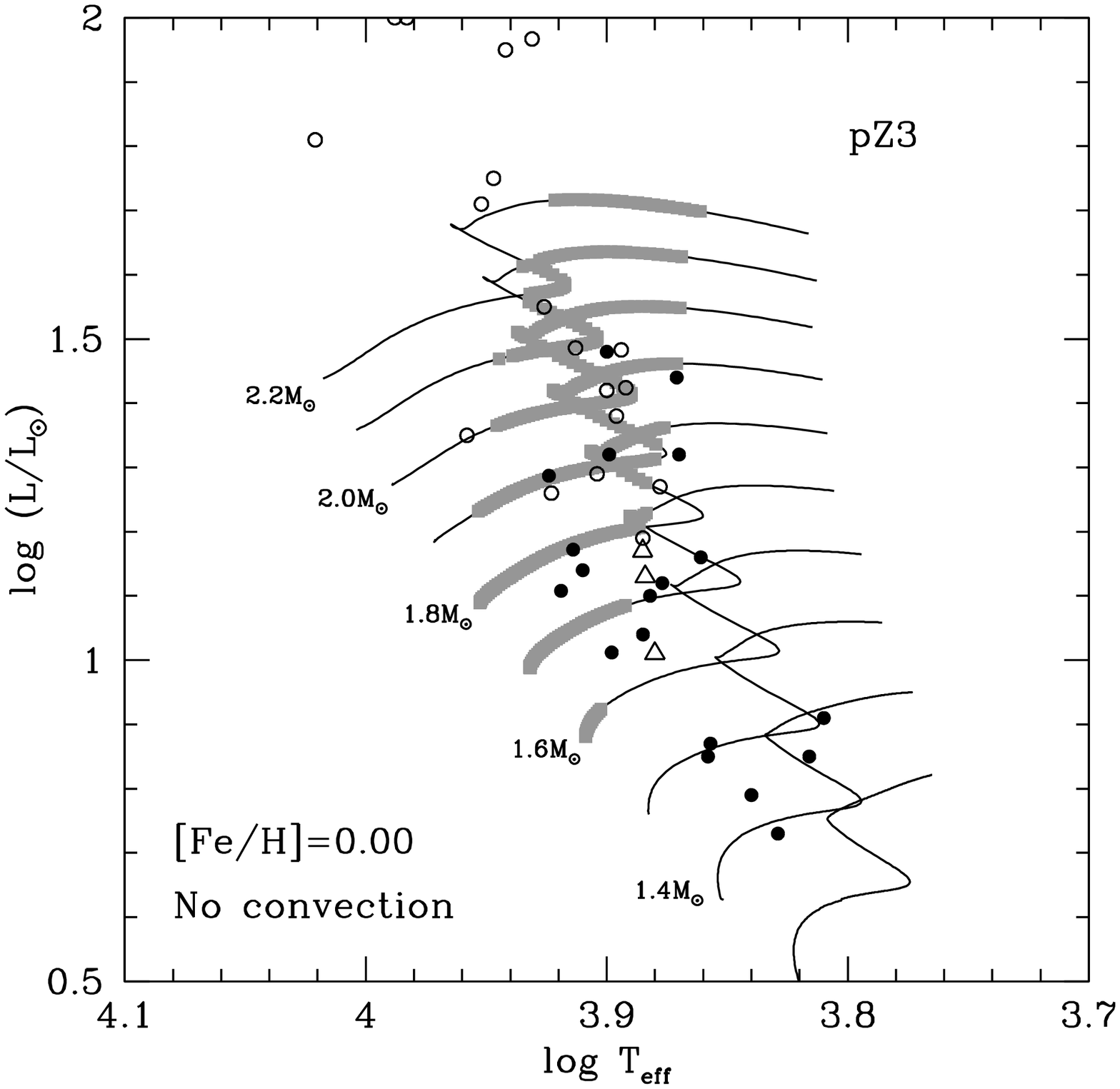}
   \includegraphics[width=0.45\textwidth]{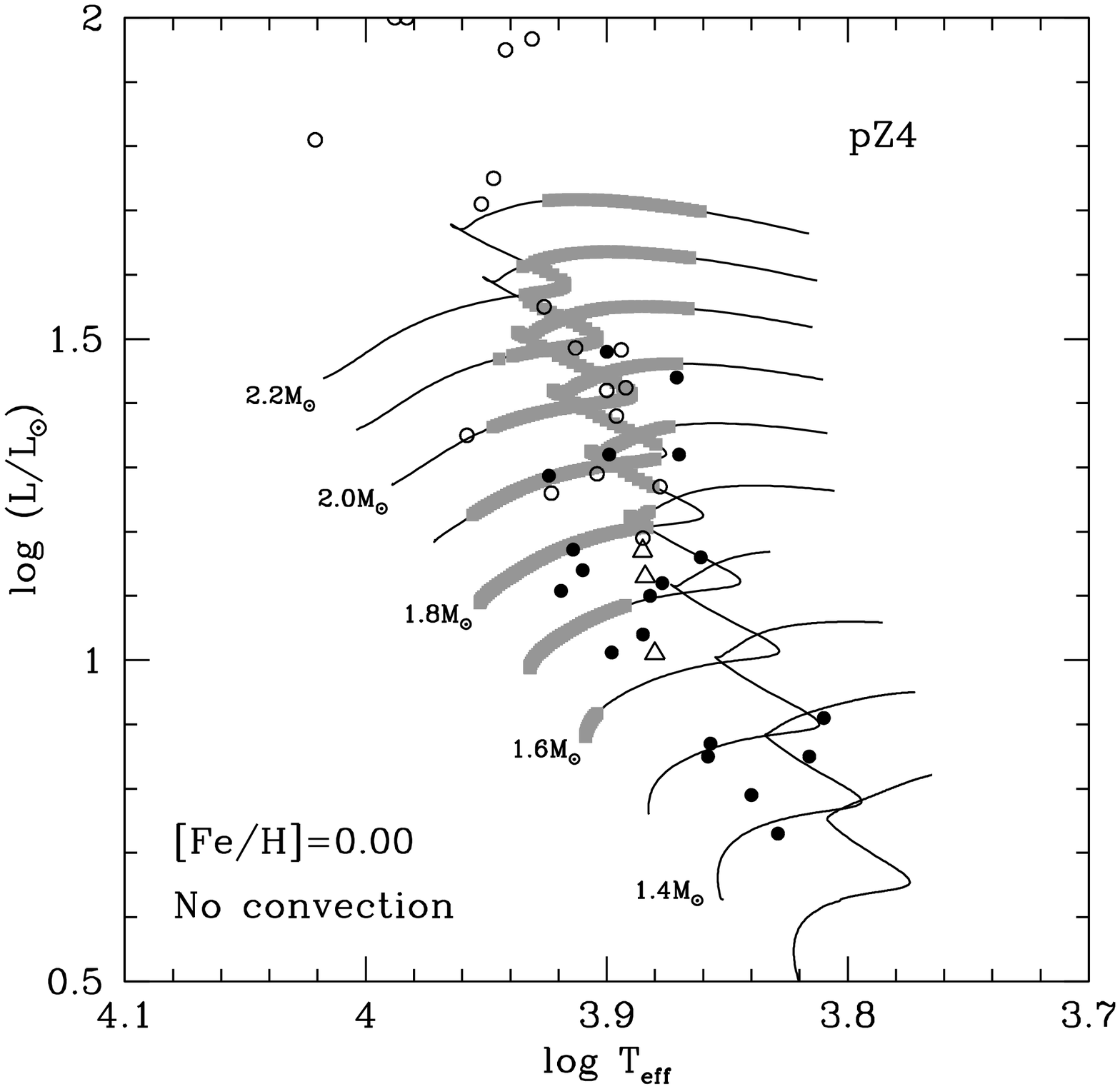}
   \includegraphics[width=0.45\textwidth]{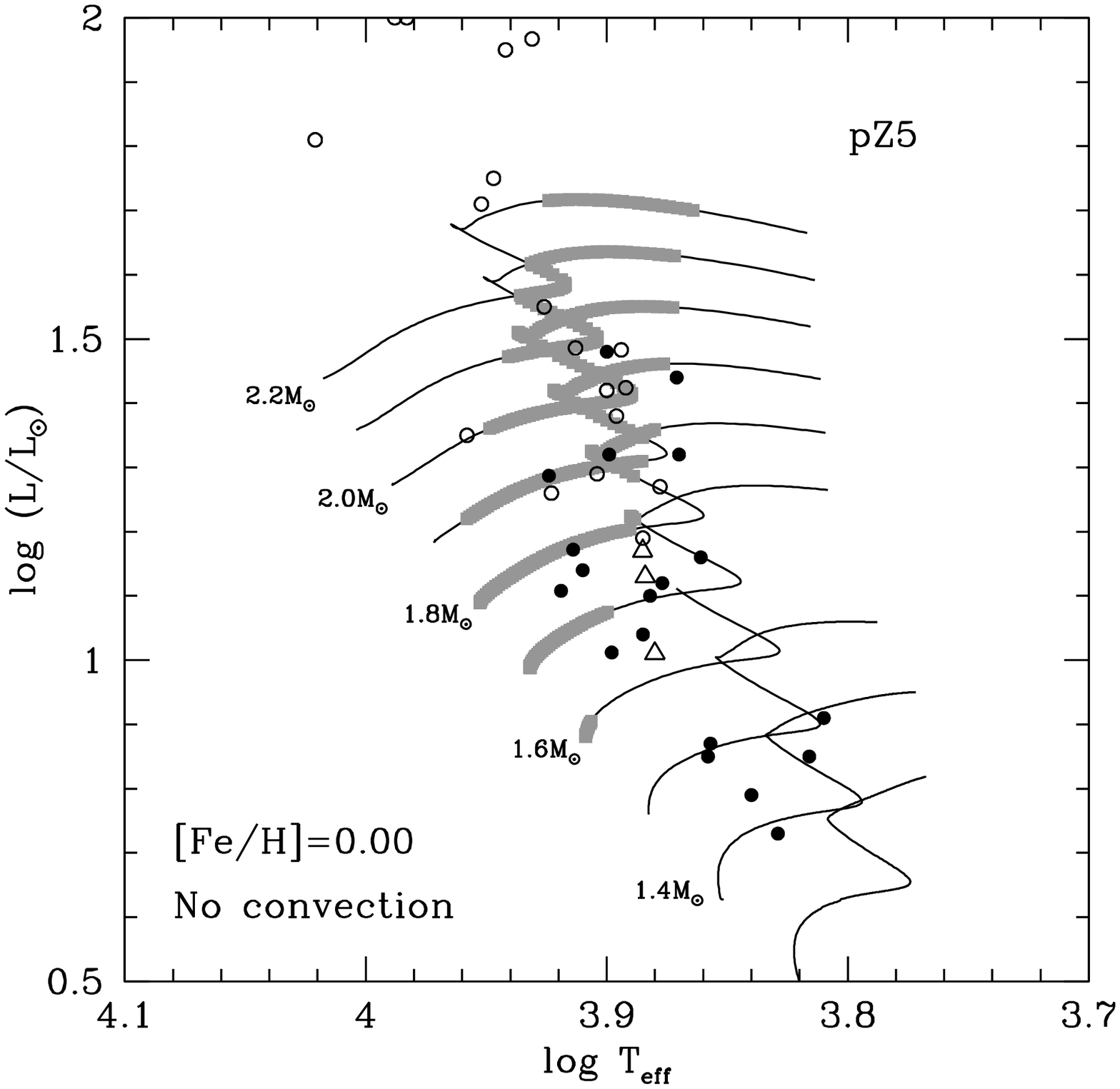}
   \includegraphics[width=0.45\textwidth]{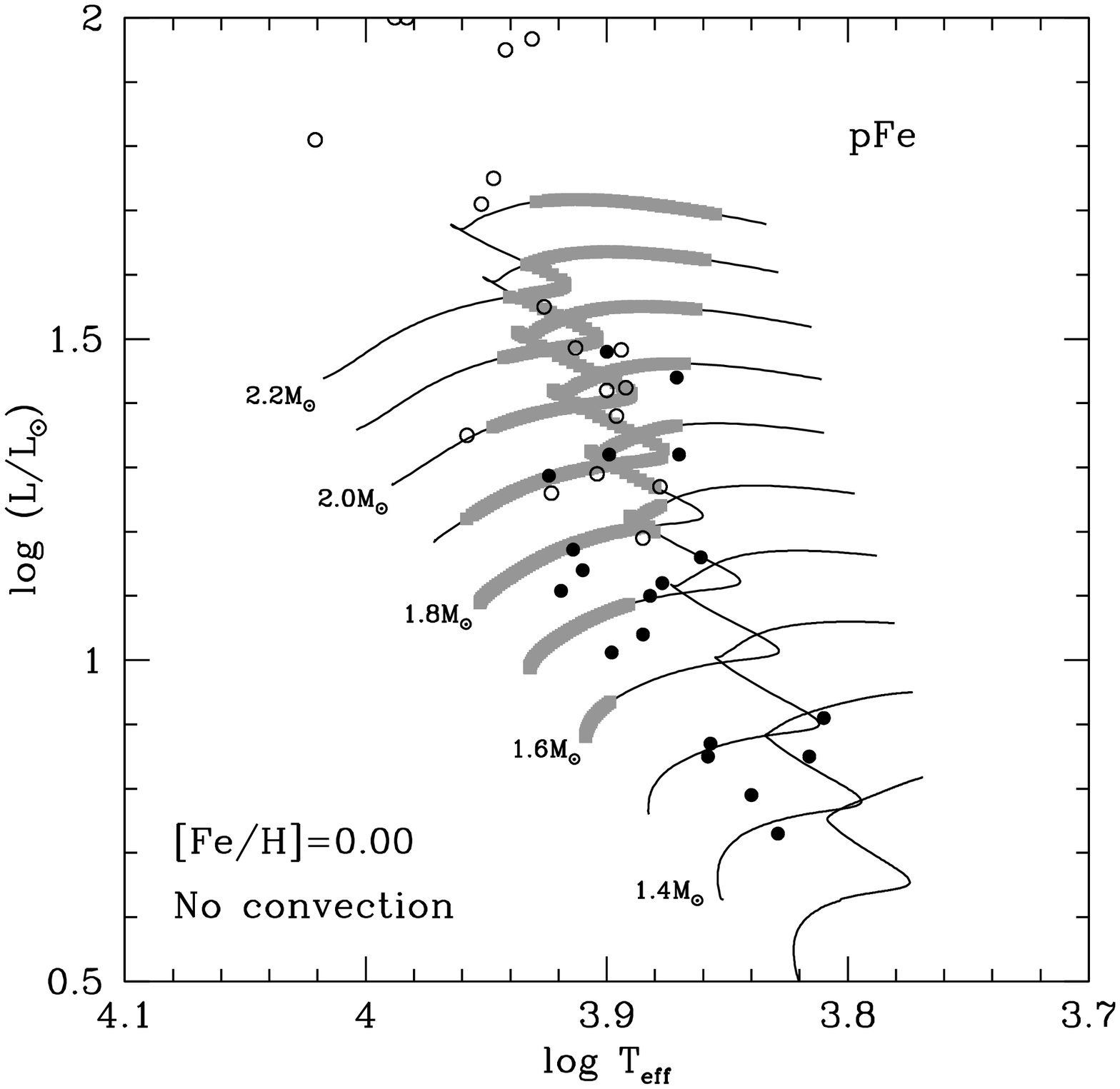}
   \caption{Figure pZ1 to pZ5 : theoretical instability strips deduced from grids of models including the metal accumulation profiles shown in Fig. \ref{profilZ}. Figure pFe : theoretical instability strip deduced from models including the iron accumulation profile presented in Fig. \ref{pfe1}. The theoretical instability strips and the observations are represented following the same conventions as in Fig. \ref{isz}.  }
              \label{diaghrpZ}%
   \end{figure*}

Figure \ref{diaghrpZ} presents the theoretical instability strips obtained for grids of models including the metal parametric accumulation profiles of Fig. \ref{profilZ}.

\begin{figure}
\includegraphics[width=0.45\textwidth]{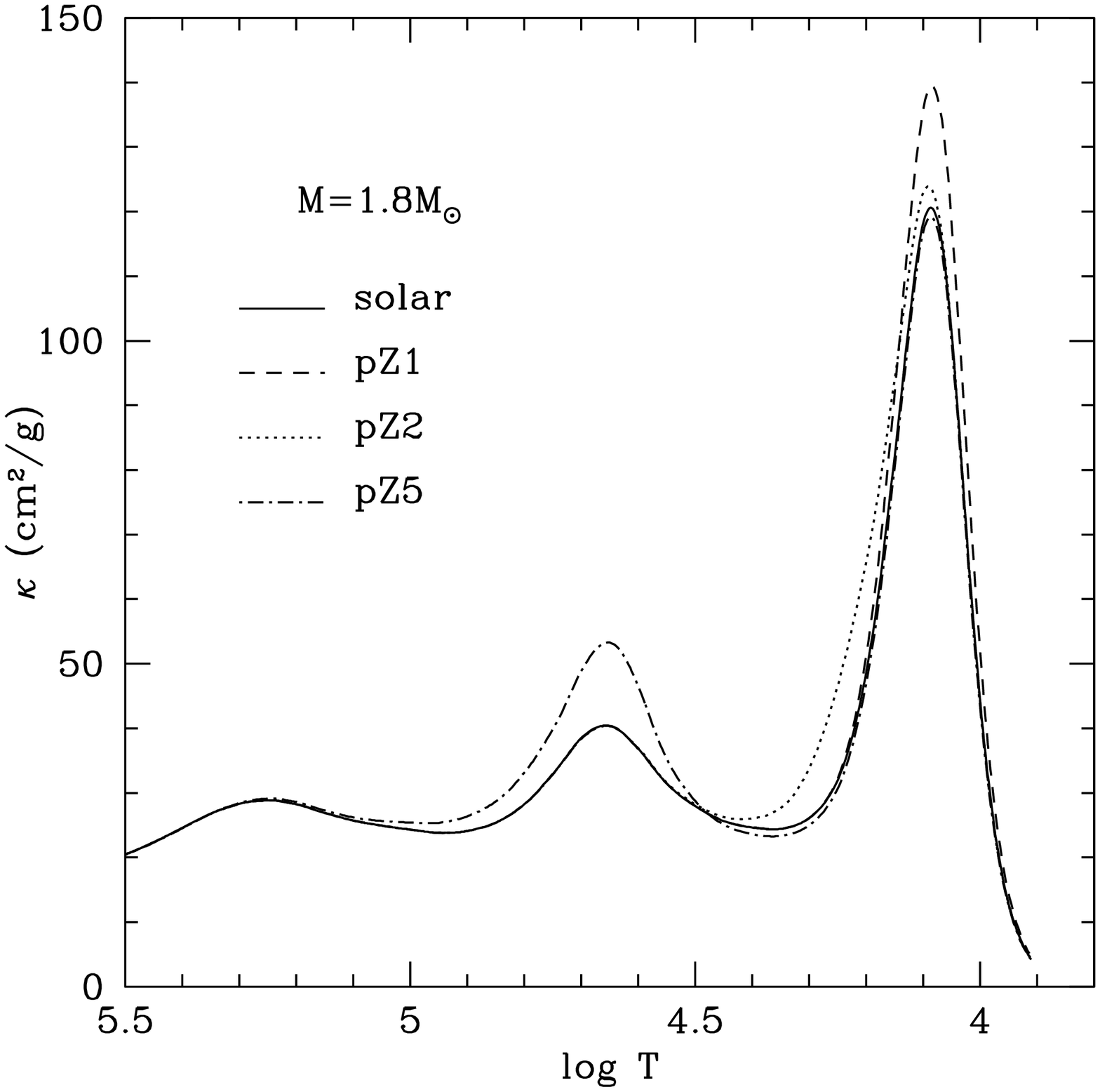}
   \caption{Opacity in models with different profiles of $Z$ (constant, pZ1, pZ2 and pZ5 according to Fig.~\ref{profilZ}) but located at the same point in the HR diagram.}
              \label{kappapz}%
   \end{figure}

\begin{figure}[h]
\resizebox{\hsize}{!}{\includegraphics{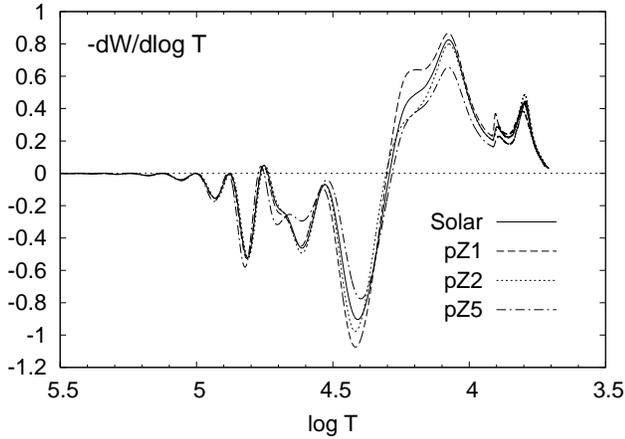}}
\caption{\label{dWroap-pZ125}Comparison of the differential work $-{\rm d}W/{\rm d}\log T$ in models with different profiles of $Z$ (constant, pZ1, pZ2 and pZ5 according to Fig.~\ref{profilZ}) but located at the same point in the HR diagram.}
\end{figure}

We see that here also, the theoretical instability strips are not affected significantly
by changes in the models. To understand more closely what is happening here, we
consider four models at the same location in the HR diagram: M = 1.8 M$_{\odot}$,
$T_{\rm eff} = 7600$~K, $\log(L/L_{\rm odot})= 1.209$, with different $Z$ profiles: constant, pZ1, pZ2, and pZ5, according to Fig.~\ref{profilZ}. These four models are located close to the red edge of the instability strip. We present in Fig.~\ref{kappapz} the opacities found for these four models. The opacity bumps are higher in the region where $Z$ is increased.
Figure \ref{dWroap-pZ125} shows the differential work found for the mode p$_{29}$ in these four models. Despite the differences in the opacity and for the same reasons discussed in Sect. \ref{rededge} and \ref{compcal}, no significant difference in the driving mechanism is found between these four models, which explains why the location of the theoretical instability strip is not
significantly affected by local changes in $Z$.

\end{appendix}

\begin{appendix}
\section{Iron local enhancement}
\label{localFe}
In this section, we investigate the effects of heavy element accumulation with a different approach. 
We study models including an iron local enhancement but a constant metal mass fraction in the radiative interior. 
These computations are motivated by results from \citet{richard01}, which demonstrated that an iron accumulation due to radiative levitation may occur in the external regions of A stars while keeping the metal mass fraction constant. 

We introduce in our models a parametric iron profile to simulate qualitatively a diffusion-induced iron accumulation. We center the iron accumulation profile on the HI ionization region where most of the driving is supposed to take place, and where iron is expected to accumulate. The metal mass fraction remains constant in the radiative interior during the entire main-sequence evolution, but the composition of the metal mixture varies with depth: the iron abundance is increased in the external layers, while the other elements are decreased proportionally, keeping the metal mass fraction Z constant. The initial hydrogen mass fraction is the same as previously (X=0.71), the metallicity is still the solar one (Z/X=0.0165, AGS05) and the models have fully radiative envelope. 

In these computations, the opacity is computed by interpolation between five tables built with different relative abundances of iron in the metal mixture (Fe$\times$0, Fe$\times$1, Fe$\times$2, Fe$\times$5, Fe$\times$10), the other elements being decreased proportionally). These tables were computed from the OPAL website; they were completed at low temperature with tables computed specifically by J. Ferguson and based upon calculations from \citet{ferguson05}.

\begin{figure}
     \includegraphics[width=0.45\textwidth]{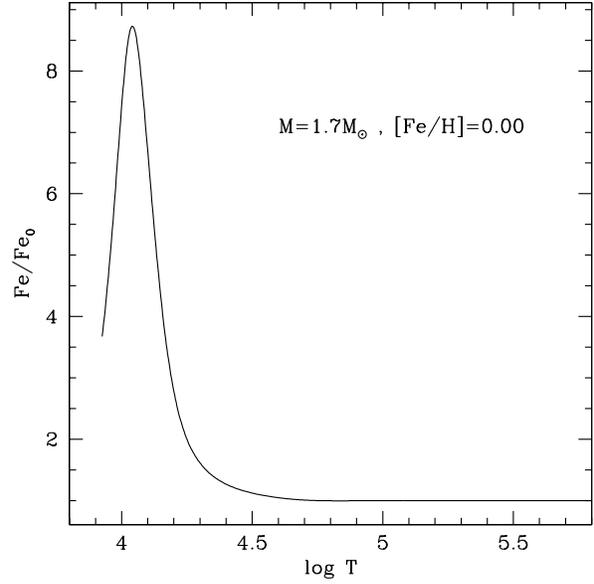}
     \caption{Iron accumulation profiles introduced in a 1.7M$_{\odot}$ model.}
     \label{pfe1}%
   \end{figure}

\begin{figure}
     \includegraphics[width=0.45\textwidth]{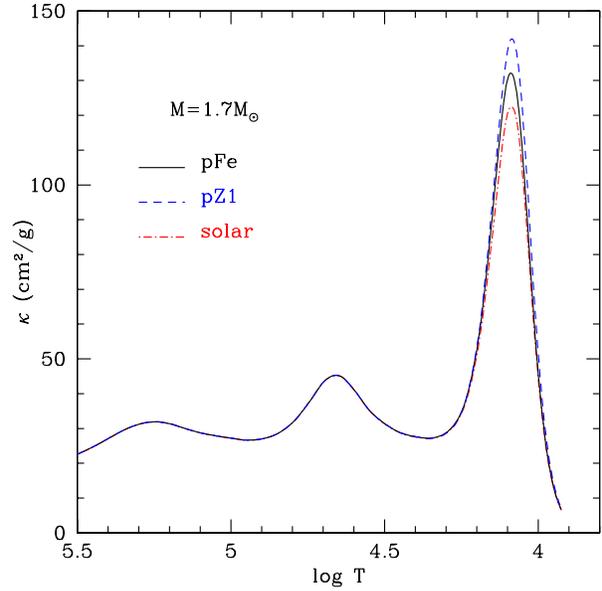}
     \caption{Opacity in a main sequence 1.7M$_{\odot}$ model (labelled ``pFe'') including the iron accumulation profile displayed in Fig. \ref{pfe1}. For comparison the opacity of two others 1.7M$_{\odot}$ models located at the same point  in the HR diagram are also shown: the 
model labelled ``solar'' includes the AGS05 solar metal mixture, the pZ1 model is the one displayed on Fig. \ref{profilZ}.} 
     \label{kappapFe}%
   \end{figure}

As an example, Fig. \ref{pfe1} displays the iron accumulation profile introduced in the 1.7M$_{\odot}$ model. The same profile is used for all the computed models with masses between 1.3 to 2.2M$_{\odot}$. 
The resulting opacity profile (referred as ``pFe'') is shown in Fig. \ref{kappapFe}. The ``solar'' and ``pZ1'' labelled curves represent the opacity profiles in two other 1.7M$_{\odot}$ models located at the same place in the HR diagram (T$_{\rm eff}$=7792K, $\rm \log$(L/L$_{\odot}$)=1.08) near the red edge of the theoretical instability strip: the ``solar'' model includes the solar metal mixture (AGS05), the pZ1-model includes the metal accumulation profile as shown in Fig. \ref{profilZ}.

Figures \ref{profilZ}, \ref{pfe1} and \ref{kappapFe} show that increasing (in an adequate way) the iron mass fraction while keeping Z constant may produce effects on the opacity similar to a local metallicity increase.
This shows that the opacity in the HI ionization region is indeed sensitive to the iron abundance, which could therefore affect the excitation mechanism of Ap stars.

The right lower HR diagram in Fig. \ref{diaghrpZ} displays the theoretical instability strip deduced from this new grid of models. This Fig. together with the middle left panel of Fig. \ref{isz} show that an iron local enhancement around logT=4.05 leads to a theoretical roAp star instability strip similar to that deduced from chemically homogenous models of solar composition.
As a consequence, diffusion-induced iron or metal accumulations in the external layers of magnetic A stars do not account for the position of the observed instability strip of roAp stars.
\end{appendix}
\end{document}